\documentclass[10pt,letterpaper]{article}

\usepackage[top=0.85in,left=2.75in,footskip=0.75in]{geometry}
\usepackage{subfigure}
\usepackage{algorithm}
\usepackage{algorithmic}
\usepackage{dsfont}
\usepackage[numbers,sort&compress]{natbib}
\usepackage{amsmath,amssymb}
\usepackage{changepage}
\usepackage[utf8x]{inputenc}
\usepackage{textcomp,marvosym}
\usepackage{cite}
\usepackage{nameref,hyperref}
\usepackage[right]{lineno}

\usepackage{microtype}
\DisableLigatures[f]{encoding = *, family = * }

\usepackage[table]{xcolor}

\newcommand{\beginresponse}[0]{\color{black}} %macro to change colour for response to reviewers/resubmission
\newcommand{\imdonenow}[0]{\color{black}} %endresponse is apparently already used as a command somewhere
\newcommand{\beginsecondresponse}[0]{\color{black}} %macro to change colour for response to reviewers/resubmission

\usepackage{array}

\newcolumntype{+}{!{\vrule width 2pt}}

\newlength\savedwidth

\usepackage{setspace} 
\raggedright
\usepackage[aboveskip=1pt,labelfont=bf,labelsep=period,justification=raggedright,singlelinecheck=off]{caption}

\makeatletter
\renewcommand{\@biblabel}[1]{\quad#1.}
\makeatother

\date{}

\usepackage{lastpage,fancyhdr,graphicx}
\usepackage{epstopdf}
\fancyheadoffset[L]{2.25in}
\fancyfootoffset[L]{2.25in}
\newcommand{\beginsupplement}{%
        \setcounter{table}{0}
        \renewcommand{\thetable}{S\arabic{table}}%
        \setcounter{figure}{1}
        \renewcommand{\thefigure}{S\arabic{figure}}%
     }

\begin{document}
\vspace*{0.2in}

\begin{flushleft}
{\Large
\textbf\newline{The impact of temporal sampling resolution on parameter inference for biological transport models} 
}
\newline
\\
Jonathan U. Harrison* \textsuperscript{1},
Ruth E. Baker\textsuperscript{1},
\\
\bigskip
\textsuperscript{\textbf{1}} Wolfson Centre for Mathematical Biology, Mathematical Institute, University of Oxford, United Kingdom
\\
\bigskip
* harrison@maths.ox.ac.uk

\end{flushleft}

% ---------------------------------------------------------------------------------------------

\section*{Abstract}
Imaging data has become an essential tool to explore key biological questions at various scales, for example the motile behaviour of bacteria or the transport of mRNA, and it has the potential to transform our understanding of important transport mechanisms. Often these imaging studies require us to compare biological species or mutants, and to do this we need to quantitatively characterise their behaviour. Mathematical models offer a quantitative description of a system that enables us to perform this comparison, but to relate mechanistic mathematical models to imaging data, we need to estimate their parameters. In this work we study how collecting data at different temporal resolutions impacts our ability to infer parameters of biological transport models; performing exact inference for simple velocity jump process models in a Bayesian framework. The question of how best to choose the frequency with which data is collected is prominent in a host of studies because the majority of imaging technologies place constraints on the frequency with which images can be taken, and the discrete nature of observations can introduce errors into parameter estimates. In this work, we mitigate such errors by formulating the velocity jump process model within a hidden states framework. This allows us to obtain estimates of the reorientation rate and noise amplitude for noisy observations of a simple velocity jump process. We demonstrate the sensitivity of these estimates to temporal variations in the sampling resolution and extent of measurement noise. We use our methodology to provide experimental guidelines for researchers aiming to characterise motile behaviour that can be described by a velocity jump process. In particular, we consider how experimental constraints resulting in a trade-off between temporal sampling resolution and observation noise may affect parameter estimates.
\beginresponse
Finally, we demonstrate the robustness of our methodology to model misspecification, and then apply our inference framework to a dataset that was generated with the aim of understanding the localization of RNA-protein complexes.
\imdonenow

% ---------------------------------------------------------------------------------------------

\section*{Author summary}
We consider how the temporal resolution of imaging studies affects our ability to carry out accurate parameter estimation for a stochastic biological transport model. This model provides a mechanistic description of motile behaviour and is often used to interrogate transport processes, such as the motion of bacteria. Parameter inference is necessary to characterise different types of transport and to make predictions about biological behaviour under different conditions. Typically, observations of the transport process, at the level of individual trajectories, are made at discrete times. This can lead to errors in parameter estimation because we do not have complete trajectory information. We present a framework for Bayesian inference for these models of biological transport processes. Using this framework, we study the effects of collecting data more or less frequently, and with varying measurement noise, on what we can learn about the biological system via parameter estimation.

% ---------------------------------------------------------------------------------------------

\section*{Introduction}

Biological transport processes occur on a wide range of spatial and temporal scales, and a common mechanism for transport involves two phases: fast active transport, and a quasi-stationary reorientation phase.
This pattern of movements has been observed at a range of scales from the intracellular transport of cellular components such as mRNA particles moving on a microtubule network \citep{parton2014subcellular},
to the run-and-tumble motion of bacteria such as \textit{Escherichia coli} \citep{othmer1988models, berg1972chemotaxis, rosser2013novel}, and the flights of birds between nesting sites \citep{taylorking2015birds}. To capture appropriately these two phases of motion, a class of models known as velocity jump process (VJP) models \citep{othmer1988models,othmer2000diffusion,codling2005calculating,codling2008random,taylorking2015birds}
(also known as correlated random walks \citep{kareiva1983analyzing,bovet1988spatial,jonsen2005robust,liepe2012calibrating,jones2015inference} or Levy Walks \citep{viswanathan1996levy,viswanathan2000levy,edwards2007revisiting,benhamou2007many}) have been developed.

Estimating the parameters of these models can give us mechanistic information relating to the underlying biological process, such as the rate of reorientation. Being able to obtain accurate estimates, with appropriate uncertainty, for these parameters allows us to compare different biological species or mutants, and gain an understanding of the underlying mechanistic behaviour. Importantly, parameterising models and quantifying the uncertainty in parameter estimates, as can be achieved via Bayesian inference, enables us to use models to make quantitive predictions of behaviour in new conditions with quantifiable uncertainty. By performing experiments to test model predictions, we can evaluate the areas in which a given model fails to describe experimental data, and so iteratively refine our understanding of a given system or phenomenon.

In this work, we consider the effects that experimental design can have on the information we can obtain from a data set, in terms of using that data to estimate parameters of a \beginresponse mechanistic \imdonenow model. 
In particular, for time series data describing a biological transport process, we vary the time between successive measurements. We demonstrate a framework for estimating the parameters of a VJP model for data of this form in the presence of noise, and examine how the posterior estimates of the model parameters change for more coarsely sampled and noisier datasets. Our framework formulates the VJP model as a process with hidden states, as in a hidden Markov model (HMM), which allows us to use particle Markov Chain Monte Carlo (pMCMC) methods to perform exact Bayesian inference. We use our framework to suggest sensible experimental design choices in the context of microscopy studies, where there may be a trade-off between how frequently it is possible to image, and the noise resulting from more or less frequent observations of the process.
We present a comparison between the pMCMC framework described in this work and approximate Bayesian computation (ABC) for parameter inference in this context. 
\beginresponse We demonstrate robustness to model mispecification in a situation where we attempt inference on synthetic data generated with a different model to the assumed model. 
Finally, we apply our method to an imaging datset of tracks of RNA-protein complexes allowing us to estimate motility parameters for a mechanistic model. \imdonenow

% ---------------------------------------------------------------------------------------------

\subsection*{Velocity jump process models}\label{VJP_intro}

VJP models have been developed to describe biological transport processes where there is persistent or biased random motion \citep{othmer1988models,painter2009modelling,taylorking2015birds}.
Correlated, persistent motion is observed experimentally in a variety of contexts \citep{mandeville1995intracellular,pankov2005rac}.  VJP models are most appropriate for motion consisting of multiple phases, such as a fast directed phase and a stationary or reorientation phase, and these models describe how an object moves in one direction before reorienting and moving in a new direction.
This type of ``run and reorientate'' motion is exhibited, for example, by \textit{Escherichia coli} \citep{berg1972chemotaxis, rosser2013novel}, fibroblasts \citep{pankov2005rac}, and RNA-protein complexes \citep{parton2014subcellular}.  

\begin{figure}[h!]
\begin{center}
\includegraphics[width=0.75\columnwidth]{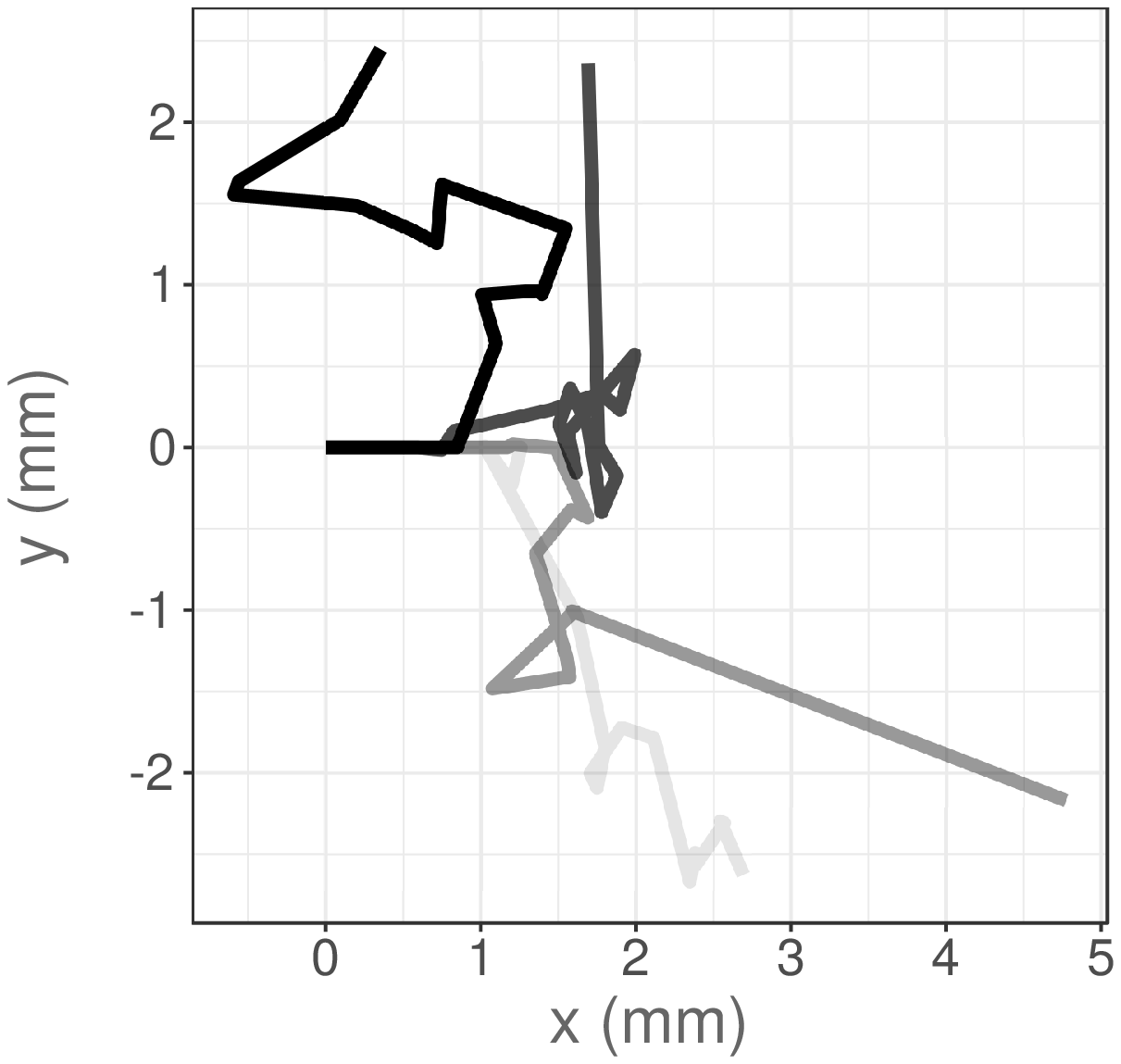}
\end{center}
\caption{Four example VJP trajectories, simulated with a uniform reorientation kernel, \beginresponse running speed $c = 50\,\mu\text{m}\text{s}^{-1}$, exponential running time distribution, $f_{\tau}(t) = \lambda e^{-\lambda t}$, with reorientation frequency $\lambda = 0.2\,\text{s}^{-1}$
and no waiting time between runs, \imdonenow for a duration of $T_{\text{final}}=64\,\text{s}$.
These trajectories start from the origin, and are orientated initially parallel to the positive $x$-axis.}
\label{Fig:VJP_example}
\end{figure}

We present here a mathematical description of a VJP model. Suppose we have a running time distribution, with probability density function (pdf) $f_{\tau}$, and a waiting time distribution, with pdf $f_{\mu}$. 
Random variables drawn from these distributions dictate the lengths of time spent in the fast active transport, or running, phase of the VJP and the reorientation phase, respectively. After the reorientation phase, a \beginresponse velocity for the new run is chosen according to a transition kernel, $f_v$.
This transition kernel could incorporate a distribution of speeds as well as directions, or could rely on a fixed speed and specify only the directionality of the run.
In the fixed speed case, 
\begin{equation*}
f_v(\mathbf{v}) = \frac{\delta_0(|\mathbf{v}| - c)}{|\mathbf{v}|}f_{\Phi}(\phi),
\end{equation*}
where $f_{\Phi}$ is a reorientation kernel descibing the angle change, and $c$ is the constant running speed. \imdonenow
For biological processes with a distinct separation of timescales between the running and reorientation phases, it is often possible to assume that reorientations between successive runs occur instantaneously and therefore neglect the reorientation phase in a model of the process \citep{othmer1988models, taylorking2015birds}.
\beginresponse Repeated simulation from this model can be used to generate trajectories of individual particles (see Fig \ref{Fig:VJP_example}, for example). \imdonenow

% ---------------------------------------------------------------------------------------------

% ---------------------------------------------------------------------------------------------

In the simplest case, where we assume that the running time distribution is memoryless, that is, exponentially distributed, with $f_{\tau} (t) = \lambda \exp (-\lambda t) $, the long time behaviour of the mean squared displacement scales linearly with time, $t$ \citep{othmer1988models}.
Further moments of the motion of individuals displaying such VJP behaviour have been characterised by making certain closure assumptions \citep{codling2005calculating}.
For the case of a more general running time distribution, with finite mean and variance, in the large time limit the probability of the particle being at position $x$ at time $t$ follows a diffusion equation \citep{taylorking2015birds}. 
 
In practice, VJP models are often parameterised by obtaining measurements of the effective diffusion coefficient or the mean squared displacement, and using these data to estimate the parameters of a specific running distribution \citep{othmer1988models,taylorking2015birds}. \citet{rosser2013novel} parameterised a HMM via maximum likelihood estimation, whilst \citet{nicosia2017general} fitted a hidden state random walk model to animal movement data using an expectation-maximization algorithm.
\beginresponse
These frequentist approaches can provide useful point estimates of the VJP parameters, and quantify the error in these estimates.
However, Bayesian approaches can propagate uncertainty from both process noise and measurement noise, which can be crucial when dealing with noisy biological data (see Fig \ref{Fig:VJP_example}).
In addition, they provide the added benefit that we can interpret the results of a Bayesian analysis as probabilistic statements about the model parameters.
%In this work we adopt a Bayesian approach to parameter estimation and obtain posterior distributions for model parameters of interest, which gives a quantifiable uncertainty to our estimates.
This quantification of uncertainty enables \imdonenow the generation of predictions of further biological behaviour using the model.
In addition, we can consider the effects of noisy data measurements upon the accuracy of the inferred parameters distributions, something which has previously been difficult to deal with in practice, or has been neglected.

For simplicity, in this work we will assume that there is a separation of timescales between \beginresponse the lengths of the \imdonenow running and reorientation phases, such that reorientations can be considered instantaneous. In addition, we assume that the running time distribution, $f_{\tau}$, is exponentially distributed, \beginresponse that particles run at a fixed speed, $c$, \imdonenow and the reorientation kernel is a uniform distribution on $[-\pi,\pi)$.
\beginresponse That is $f_{\tau} (t) = \lambda \exp (-\lambda t) $ and $f_{\Phi}(\phi) = 1/2\pi \,\mathds{1}_{[-\pi,\pi)}(\phi)$. \imdonenow 
We follow the trajectory of a single, motile individual, and take, as experimental measurements, the change in angle between successive observed positions (calculated relative to the previous observed position, \beginresponse see Fig \ref{Fig:set_up} \imdonenow ), subject to measurement noise drawn from a wrapped Normal distribution, $N(0,\sigma^2)$, where $\sigma$ is the magnitude of the noise. % (see Fig \ref{Fig:set_up} where the observed angle change between observations at times $k\Delta{t}$ and $(k+1)\Delta{t}$ is $\theta_1$).
Our Bayesian inference approach will target estimation of the reorientation rate, $\lambda$, and the magnitude of the noise, $\sigma$.

% ---------------------------------------------------------------------------------------------

\subsection*{Inference for velocity jump process models via particle Markov chain Monte Carlo}\label{pmcmc_intro}

Inference for partially observed Markov processes can be performed using particle Markov Chain Monte Carlo (pMCMC), as developed by \citet{andrieu2009pseudo} and \citet{andrieu2010particle}. pMCMC provides a Bayesian framework for parameter estimation by allowing samples to be drawn from the posterior distribution of the model parameters, given observed data, without needing to evaluate the likelihood function directly. For partially observed Markov process models, the model structure makes directly evaluating the likelihood difficult or expensive\footnote{We note that VJP models can be viewed in this form by introducing hidden states, as explained in Section `Methods'.}. Instead of evaluating the likelihood directly, we can use (unbiased) estimates of the likelihood within an MCMC algorithm. Estimating the likelihood of the observed data given certain parameters can be achieved with a particle filter (also known as a sequential Monte Carlo scheme) \citep{gordon1993novel} for a fixed, finite, number of particles. The results of \citet{andrieu2010particle} demonstrate that, even when a finite number of particles are used in the filter to estimate the likelihood, the MCMC algorithm will still target the correct posterior distribution.
These methods have been applied in the context of modelling epidemics \citep{rasmussen2011inference} and biochemical reaction networks \citep{golightly2011bayesian,golightly2015delayed}.
However, pMCMC methods for parameter inference have not previously been applied to spatial agent-based models, such as the VJP model considered here.
\beginresponse
The details of the pMCMC algorithm used is this work are given in Section `Methods'.
\imdonenow
% ---------------------------------------------------------------------------------------------

\section*{Methods}

\begin{figure}[h!]
\begin{center}
\includegraphics[width=0.75\columnwidth]{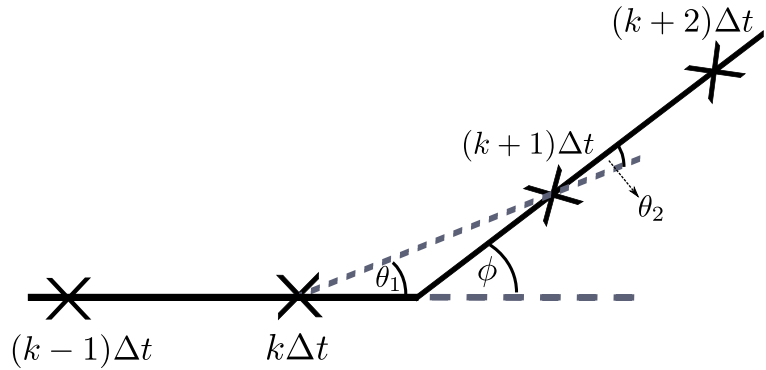}
\end{center}
\caption{The motile individual is observed at discrete times $t=k\Delta t$ for $k\in\{0,1,\ldots,T\}$ (marked by crosses). In each time interval between observations, we measure the angle change between the observed direction of travel between successive observations. We assign a hidden state to each time interval according to whether a reorientation event occurred in that time interval or not. In this example, which excludes measurement noise, there is a reorientation, through angle $\phi$, between observations at times $t=k\Delta{t}$ and $t=(k+1)\Delta{t}$; however, the measured angle change between these observations is $\theta_1$. Conversely, there is no reorientation event between observations at $t=(k+1)\Delta{t}$ and $t=(k+2)\Delta{t}$ and an observed angle change of $\theta_2$.
}
\label{Fig:set_up}
\end{figure}

We will demonstrate how to exploit pMCMC methods to obtain posterior parameter estimates for the VJP outlined in the Introduction by formulating the model within an appropriate framework that incorporates hidden states. This formulation additionally allows us to incorporate a model for measurement noise in our observations, as well as explicitly accounting for the temporal discretisation of the data.
The hidden states (also known as latent variables) in our model will describe whether or not a reorientation event occurred between observations of the VJP. This is not a variable that we can observe directly, since we only measure the observed angle change. For example, in Fig~\ref{Fig:set_up} there was a reorientation event between observations at $t=k\Delta{t}$ and $t=(k+1)\Delta{t}$, and an observed angle change of $\theta_1$.
On the other hand, there was no reorientation event between observations at $t=(k+1)\Delta{t}$ and $t=(k+2)\Delta{t}$ but still a non-zero observed angle change of $\theta_2$. Note that the true reorientation angle for the reorientation event that takes place between observations at $t=k\Delta{t}$ and $t=(k+1)\Delta{t}$ is $\phi$.

% ---------------------------------------------------------------------------------------------

% ---------------------------------------------------------------------------------------------

\begin{figure}[h!]
\begin{center}
\includegraphics[width=0.75\columnwidth]{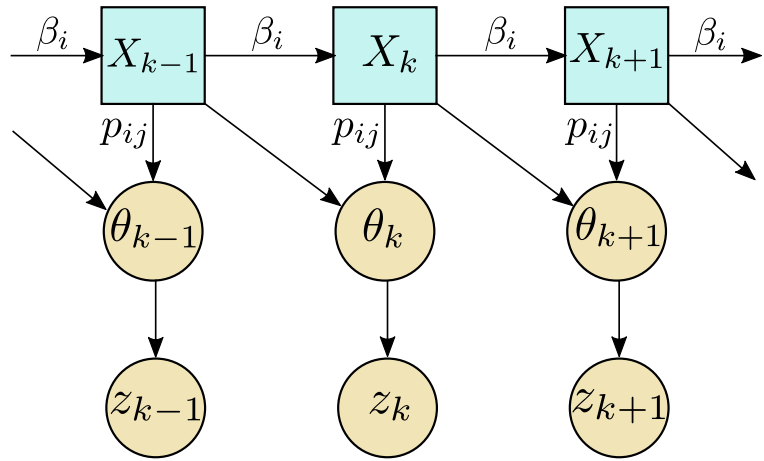}
\end{center}
\caption{Hidden and observed states in a partially observed Markov process model.
\beginresponse
We observe an angle change, $\theta_k$, which is dependent on a hidden state $X_k$ defined in Eq. \eqref{hidden_state_xk}.
This dependency is shown by the arrows between states. 
Here, we require dependencies on the previous hidden state, $X_{k-1}$, since observation times will not coincide with reorientation events in general.
Additionally, we introduce an extra layer of state dependencies to capture the measurement noise in this noisy biological system.
We suppose that $\theta_k$ is the true observed angle change and our noisy observed version of this is $z_k$, with dependencies shown by the arrows.
\imdonenow
}
\label{Fig:Basic_hmm}
\end{figure}

We assume that we observe the system at times $\{k \Delta t$ : $k\in\{0,1,\ldots,T\} \}$, up until a final time $T_{\text{final}}=T\Delta t$, by measuring the position of the individual of interest and recording the observed angle change. We define the hidden variable, $X_k$, as follows: 
\begin{equation} \label{hidden_state_xk}
X_k = \begin{cases}
       1, \hspace{5mm} \text{ if a reorientation event occurred during } [k\Delta t, (k+1)\Delta t), \\
       0, \hspace{5mm} \text{ otherwise.}
      \end{cases}  
\end{equation}
The observed state is the observed angle change, $\theta_k$, obtained as the difference between the observed direction of travel during $[(k-1)\Delta t, k\Delta t)$ and $[k\Delta t, (k+1)\Delta t)$, $k \ge 1$, \beginresponse as illustrated in Fig \ref{Fig:set_up}.
For example, if we directly observe as data the positions $\{ (x_k,y_k)$ :  $k\in\{0,1,\ldots,T\} \}$, then we can calculate the observed angle change as
$$ \left\{ \theta_k = \text{arctan}\left(\frac{y_{k+1} - y_k}{x_{k+1} - x_{k}} \right) : k\in\{0,1,\ldots,T-1\} \right\}. $$ \imdonenow
We \beginresponse illustrate the sequence of \imdonenow hidden and observed states in Fig \ref{Fig:Basic_hmm}, with dependence between these states given by the transition and emission probabilities, denoted $\beta_i$ and $p_{i j}$, respectively. The hidden variable, $X_k$, evolves according to the VJP model.
Here, since we have an exponential distribution for the running time, \beginresponse $P(\text{reorientation event in } [(k-1)\Delta t, k\Delta t)) = 1 - \exp{(-\lambda \Delta t)}$. \imdonenow
% ---------------------------------------------------------------------------------------------

\beginresponse We assume that reorientation events are rare relative to the sampling rate such that $\lambda \Delta t \ll 1$.
This assumption allows us to neglect multiple reorientations in a single time interval, enabling us to greatly simplify the problem, so that it is tractable via the binary hidden variables $X_k$. 
\imdonenow
\beginsecondresponse
We highlight that the model structure as shown in Fig. 3 is an approximation. In reality, observed data will not be a function of only the current and previous hidden states, $X_k$ and $X_{k-1}$, but may depend on other previous states also.
The assumption that $\lambda \Delta t \ll 1$ enables us to neglect dependence on earlier states.
\imdonenow

% ---------------------------------------------------------------------------------------------

\subsection*{Model without measurement noise} \label{section:no_noise}

\begin{figure}[h!]
\begin{center}
\includegraphics[width=\columnwidth]{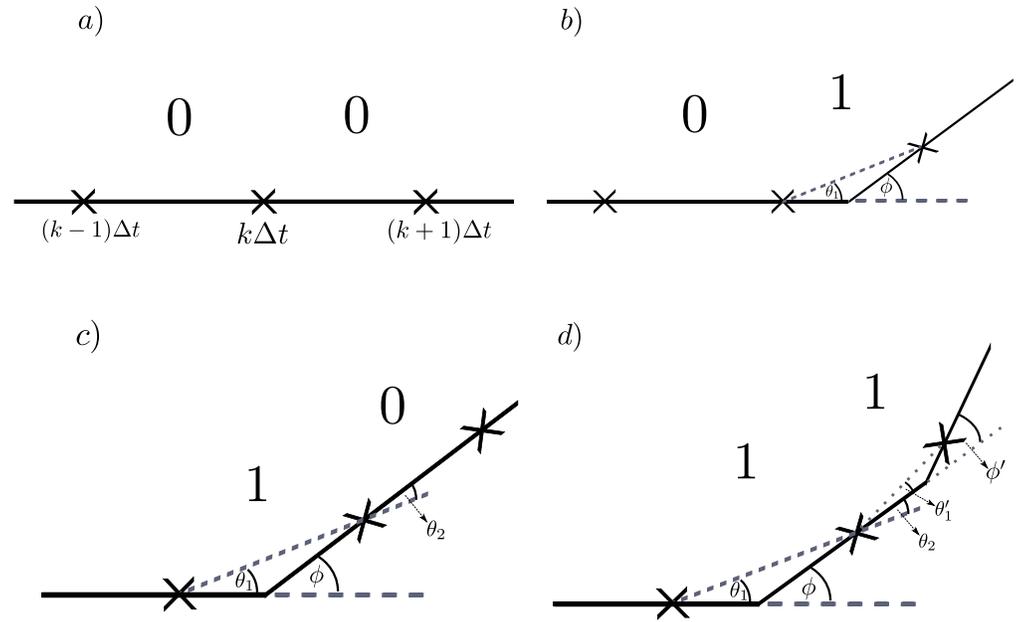}
\end{center}
\caption{Example paths for each pattern of hidden states.
The particle of interest moves from left to right, and is observed at times $(k-1)\Delta t, k\Delta t$, and $(k+1)\Delta t$.
In a), there are no reorientation events, so the particle continues along a straight trajectory.
In b), there is a single reorientation in the time interval $[k\Delta t, (k+1)\Delta t)$, where the particle turns through an angle $\phi$,
but the observed angle change is $\theta_1$ due to the discrete nature of the observations.
In c), there is a single reorientation in the time interval $[(k-1)\Delta t, k\Delta t)$.
The particle turns through an angle $\phi$ and continues on its new trajectory. 
We observe an angle change $\theta_1$ for the time interval $[(k-1)\Delta t, k\Delta t)$, and observe an angle change $\theta_2$ for the next time interval $[k\Delta t, (k+1)\Delta t)$
even though there was no reorientation during this time interval.
Similarly, in d), there was a reorientation of true angle change $\phi$ during $[(k-1)\Delta t, k\Delta t)$
followed by another reorientation of true angle change $\phi'$
in the time interval $[k\Delta t, (k+1)\Delta t)$. 
In this case, we observe angle change $\theta_1$ for the time interval $[(k-1)\Delta t, k\Delta t)$, and observe an angle change of $\theta_2 + \theta_1'$
for the time interval $[k\Delta t, (k+1)\Delta t)$.
}
\label{Fig:hidden_states}
\end{figure}

We initially make progress by simplifying the problem and its exposition via the assumption of zero measurement error. To relate the unobserved hidden state to the observed angle change, which we can measure, we derive probability distributions for the angle change given the hidden state. Note that there is additional dependence not only on the current hidden state, but also on the previous hidden states, as shown in Fig. \ref{Fig:Basic_hmm}. We can obtain expressions for the emission probabilities by considering the path that is taken under different sequences of hidden states (Fig \ref{Fig:hidden_states}); we will outline simplifying assumptions as they are made.

% ---------------------------------------------------------------------------------------------

% ---------------------------------------------------------------------------------------------

Suppose we observe the system at times $(k-1)\Delta t, k\Delta t$, and $(k+1)\Delta t$ for some $k$, as shown in Fig. \ref{Fig:hidden_states}.
The sequence of hidden states $i, j$, corresponds to whether there were reorientation events in the time intervals $[(k-1)\Delta t, k\Delta t)$ and $[k\Delta t, (k+1)\Delta t)$, respectively.
Let $p_{i j} (\theta) $ be the pdf of observing a reorientation angle of $\theta$ given the sequence of hidden states $i, j$.
We assume angle change $\theta$ was observed over the time interval $[k\Delta t, (k+1)\Delta t)$ corresponding to the hidden state $j$.
In the case where no reorientation occurs in either of the time intervals (corresponding to the situation in Fig \ref{Fig:hidden_states}a)), then, assuming no noise, we would observe zero angle change. That is, we have $p_{00} (\theta) = \beginresponse \delta_0(\theta)\imdonenow $. If a reorientation occurs in the time interval $[k\Delta t, (k+1)\Delta t)$, but not in the preceding time interval, $[(k-1)\Delta t, k\Delta t)$, (as shown in Fig \ref{Fig:hidden_states}b)), then the observed reorientation angle is $\theta_1$, as labelled in Fig \ref{Fig:hidden_states}b). This gives $p_{0 1}(\theta) := p_{\Theta_1} (\theta)$. The marginal distribution of $\Theta_1$ is derived in the Section `Derivation of emission probabilities', and depends on the running time distribution and the reorientation kernel.

If, immediately after a reorientation, we have no reorientation in the following time interval, then we may still observe a nonzero angle change since our discrete observation times do not in general coincide with the reorientation events. In this case, we have a reorientation event during $[(k-1)\Delta t, k\Delta t)$, and no reorientation during $[k\Delta t, (k+1)\Delta t)$. Such a situation with a pattern of hidden states $1, 0$ is shown in Fig \ref{Fig:hidden_states}c), and the observed angle change during the time interval $[k\Delta t, (k+1)\Delta t)$ corresponds to $\theta_2$ in the diagram. We note that, by geometric arguments, we have 
\begin{equation}           
\label{eqn:symmetry}                                                      
\theta_1 + \theta_2 = \phi, 
\end{equation}
where $\phi$ is the true angle change. Therefore, given a pattern of hidden states $1, 0$ the angles $\theta_1$ and $\theta_2$ are not independent, and we have $p_{1 0} (\theta_1, \theta_2) = p_{\Theta_2 | \Theta_1 = \theta_1} (\theta_2)$. Note that throughout we will assume that we can observe the previous angle change directly, which is equivalent to assuming that the pattern of hidden states is in fact $0, 1, 0$.

The remaining cases involve reorientation events in successive time intervals. Suppose we have the case where we have two successive reorientation events, giving a pattern of hidden states $1, 1$, which is shown graphically in Fig \ref{Fig:hidden_states}d).
This case is similar to the case with hidden states $1, 0$ (see Fig \ref{Fig:hidden_states}c)),
in that we observe a contribution to the reorientation angle, $\theta_2$, from the correction for the previous reorientation event, and also a contribution from the new reorientation event, $\theta'_1$. 
As can be seen in Fig \ref{Fig:hidden_states}d), these contributions sum to give an observed angle change for the time interval $[k\Delta t, (k+1)\Delta t)$ of $\theta = \theta_2 + \theta'_1$.
The probability density for sums of random variables can be expressed as a convolution \citep{grimmett2001probability}.
Hence, $p_{1 1} (\theta_1, \theta) = (p_{\Theta_1} \ast p_{\Theta_2 | \Theta_1 = \theta_1})(\theta)$, where $\ast$ is the convolution operator. Any further cases involve multiple reorientation events within a single time interval which occurs rarely, with probability on the order of $\mathcal{O}((\lambda\Delta t)^2)$. 
We can \beginresponse safely \imdonenow neglect these provided $\lambda\Delta t \ll 1$.

To summarise therefore, the emission probabilities (that is the probability of observing a certain angle change given the sequence of hidden states) in the case without noise can be given as follows:
\begin{align*}
p_{00} (\theta) &= \beginresponse \delta_0 (\theta); \imdonenow \\
p_{01} (\theta) &= p_{\Theta_1} (\theta); \\
p_{10} (\theta) &= p_{\Theta_2 | \Theta_1} (\theta); \\
p_{11} (\theta) &= (p_{\Theta_2 | \Theta_1} \ast p_{\Theta'}) (\theta). \\
\end{align*}

% ---------------------------------------------------------------------------------------------

\subsection*{Model with measurement noise} \label{section:noise}

To account for noise, we introduce another layer of states in our diagram of state dependencies, as shown in Fig \ref{Fig:Basic_hmm}.  Let $z_k$ be the noisy observed angle change at time $k\Delta t$, which is a noisy observation of $\theta_k$, such that for a noise model $K$ we have $z_k \sim K(\cdot,\theta_k)$. Alternatively, we can write the noisy observed angle as 
\begin{equation} \label{eq:noise}
z_k = \theta_k + \epsilon_k ,
\end{equation}
where $\epsilon_k \sim K(\cdot,0)$.
Under the assumption of a wrapped normal noise model, $\epsilon_k \sim N(0,\sigma^2)$, \beginresponse where $\sigma$ is the magnitude of the measurement noise. \imdonenow
Since Eq \eqref{eq:noise} represents $z_k$ as a sum of random variables, it becomes clear that we can obtain the noisy emission probabilities, $q_{ij}$, corresponding to a pattern of hidden states $i, j$ via the following convolution:
\begin{align*}
  q_{ij} (\theta) &= (p_{ij} \ast K) (\theta) \\
		  &= \int_{\Theta} p_{ij}(\theta - x) K(x, 0) \,\text{d} x .
\end{align*}
To extend our previous results to the noisy case, we therefore need to compute these convolutions, a task that must be carried out numerically.

% ---------------------------------------------------------------------------------------------

\subsection*{Derivation of emission probabilities}

\begin{figure}[h!]
\begin{center}
\includegraphics[width=0.5\columnwidth]{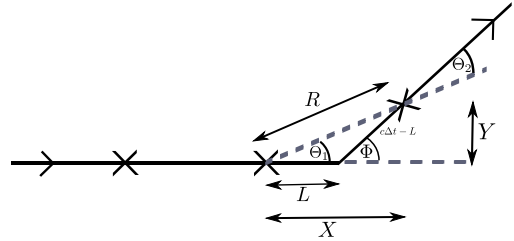}
\end{center}
\caption{True and observed angle changes \beginresponse ($\Phi$ and $\Theta_1$, respectively) \imdonenow based on discrete time observations of a VJP without measurement error. Cartesian co-ordinates $X$ and $Y$ are shown in addition to polar co-ordinates $R$ and $\Theta_1$.
The particle is moving along the trajectory from left to right as shown by the arrows.}
\label{Fig:labelled_diagram}
\end{figure}

In previous work \citep{rosser2012mathematical}, marginal distributions for the observed angle change, $\Theta_1$, were obtained and we summarise the arguments here. Fig \ref{Fig:labelled_diagram} shows the true angle change, $\Phi$, and the observed angle change, $\Theta_1$, based on discrete time observations of the VJP \beginresponse for the case of hidden states 0, 1. \imdonenow

% ---------------------------------------------------------------------------------------------

% ---------------------------------------------------------------------------------------------

By changing co-ordinates from the displacement, $L$, (which is given by the running time distribution $f_{\tau}$) and true angle change, $\Phi$, to Cartesian co-ordinates $X$ and $Y$, and then to polar co-ordinates, $R$ and $\Theta_1$, we can obtain a joint distribution for $R$ and $\Theta_1$ of 
\begin{equation*}
f_{R,\Theta_1} (r,\theta_1) = \begin{cases} \cfrac{r}{c\Delta t (c\Delta t - r\cos \theta_1) }f_{\Phi} \left( \text{arccos} \left[\frac{r^2 \sin^2 \theta_1 - (c \Delta t - r\cos \theta_1)^2}{r^2 \sin^2 \theta_1 + (c \Delta t - r\cos \theta_1)^2} \right] \right), 
\text{if } (r,\theta_1) \in C, \\
0, \hspace{3.5cm} \text{otherwise,}      
\end{cases}
\end{equation*}
where $C= \{ r \le c\Delta t, \theta \in [-\pi,\pi) \} $ is the set of permissible values for $r$ and $\theta$, $\Delta t$ is the discretisation in time
and $c$ is the running speed.
Integrating $f_{R,\Theta_1} (r,\theta_1)$ over $r$ allows us to obtain the marginal distribution for $\theta_1$ via 
\begin{equation*}
f_{\Theta_1} (\theta_1) = \int_{0}^{\infty} f_{R,\Theta_1} (r,\theta_1) \,\text{d} r. \end{equation*}
In the case where we assume a uniform reorientation kernel, $f_{\Phi}(\phi) = 1/2\pi \, \mathds{1}_{[-\pi,\pi)}(\phi)$, then this integral becomes 
\begin{align*}
 f_{\Theta_1} (\theta_1) &= \int_0^{c \Delta t} \frac{r}{2 \pi c \Delta t (c\Delta t - r\cos \theta_1)} \,\text{d} r \\
    &= \frac{-1}{2 \pi c \Delta t \cos(\theta_1)} \int_0^{c \Delta t} \left( 1 + \frac{ \frac{c \Delta t}{\cos{\theta_1}}}{r - \frac{c \Delta t}{\cos \theta_1}} \right) \text{d} r \\
    &= \frac{-1}{2 \pi c \Delta t \cos(\theta_1)} \left[ r + \frac{c \Delta t }{\cos \theta_1} \log \left\lvert r - \frac{c \Delta t }{\cos \theta_1} \right\rvert \right]_0^{c\Delta t}\\
    &= \frac{-1}{2\pi \cos^2 \theta_1} \left( \cos \theta_1 + \log (1 - \cos \theta_1) \right).
\end{align*}
We note that the marginal distribution for $\theta_1$ does not depend on the speed, $c$, or the time discretisation, $\Delta t$. This is intuitive because $f_{\Theta_1}$ is the pdf of a certain observed angle change, $\theta_1$, given there was a reorientation in that interval, irrespective of the length of that interval. 

% ---------------------------------------------------------------------------------------------

\subsubsection*{Derivation of the joint distribution of $\Theta_1$ and $\Theta_2$} \label{derivation}

By considering the displacement in the $X$ and $Y$ directions during a time step, we have
\begin{align*}
 r \cos (\theta_1 ) &= l + (c\Delta t - l) \cos (\phi ), \\
 r \sin (\theta_1 ) &= (c\Delta t - l) \sin (\phi ).
\end{align*}
Dividing these expressions, we can relate $\theta_1$ to $l$ and $\phi$ by
\begin{equation}
 \tan(\theta_1 ) = \frac{(c\Delta t - l) \sin (\phi )}{l + (c\Delta t -l) \cos (\phi )},
\end{equation}
and rearranging for $l$, we have
\begin{equation*}
l = \frac{c\Delta t (\tan \phi - \tan \theta_1 )}{\tan \phi + \tan \theta_1 (\sec \phi-1 )}.
\end{equation*}
Differentiating with respect to $\theta_1$, we obtain
\begin{equation*}
 \frac{\partial l}{\partial \theta_1} = \frac{ - c\Delta t \sec^2 \theta_1 }{\tan \phi + \tan \theta_1 (\sec \phi-1 )} - \frac{c \Delta t (\tan \phi - \tan \theta_1 ) \sec^2 \theta_1 (\sec \phi -1)}{(\tan \phi + \tan \theta_1 (\sec \phi-1 ))^2},
\end{equation*}
which can be simplified to give
\begin{equation*}
 \frac{\partial l}{\partial \theta_1} = \frac{-c\Delta t \sin \phi}{(\sin \phi \cos \theta_1 + \sin \theta_1 (1 - \cos \phi ))^2}.
\end{equation*}
To transform from coordinates $(L,\Phi)$ to coordinates $(\Theta_1,\Phi)$, we can use the Jacobian $J_{L,\Phi}$, where 
\begin{equation}
\text{det} \hspace{1mm} J_{L,\Phi} =  \frac{\partial l}{\partial \theta_1} = \frac{-c\Delta t \sin \phi}{(\sin \phi \cos \theta_1 + \sin \theta_1 (1 - \cos \phi ))^2}.
\end{equation}
Therefore the joint distribution of $\theta_1$ and $\phi$ is
\begin{equation*}
f_{\Theta_1,\Phi}(\theta_1,\phi) = \begin{cases}  
\cfrac{|c\Delta t \sin \phi |}{(\sin \phi \cos \theta_1 + \sin \theta_1 (1 - \cos \phi ))^2} . f_{\phi} (\phi) .\cfrac{1}{c \Delta t}, \hspace{3mm} \text{if } \theta_1, \phi \in C, \\
0, \hspace{3.5cm} \text{otherwise,}                                   
                                  \end{cases}
\end{equation*}
where $C= \{ (\theta_1,\phi) : \phi \in [-\pi,\pi), \theta \in (\min(0,\phi),\max(0,\phi)) \} $.
Under the assumption that $\phi$ is uniform on $[-\pi,\pi)$, as described in the Section `Velocity jump process models', such that $f_{\Phi}(\phi) = 1/2\pi \, \mathds{1}_{[-\pi,\pi)}(\phi)$, we have
\begin{equation}
f_{\Theta_1,\Phi}(\theta_1,\phi) = \begin{cases}  
\cfrac{|\sin \phi |}{2 \pi (\sin \phi \cos \theta_1 + \sin \theta_1 (1 - \cos \phi ))^2}, \hspace{3mm} \text{if } \theta_1, \phi \in C, \\
0, \hspace{3.5cm} \text{otherwise.}                                   
                                  \end{cases}
\end{equation}
Changing variables again to $(\theta_1, \theta_2)$, via $\phi = \theta_1 + \theta_2$, we have
\begin{equation}
f_{\Theta_1,\Theta_2}(\theta_1,\theta_2) = \begin{cases}  
\cfrac{|\sin (\theta_1 + \theta_2) |}{2 \pi (\sin (\theta_1 + \theta_2) \cos \theta_1 + \sin \theta_1 (1 - \cos (\theta_1 + \theta_2) ))^2}, \hspace{3mm} \text{if } \theta_1, \theta_2 \in C, \\
0, \hspace{3.5cm} \text{otherwise,}                                   
                                  \end{cases}
\end{equation}
where $C= \{ (\theta_1,\theta_2) : \theta_1 + \theta_2 \in [-\pi,\pi) \} $.
From this joint distribution, we can then find the conditional distribution of $\theta_2$ given $\theta_1$, which is required to give
$p_{10} (\theta) = f_{\Theta_1, \Theta_2}(\theta_1,\theta) / f_{\Theta_1}(\theta_1)$.
% ---------------------------------------------------------------------------------------------

\beginresponse
\subsection*{Particle MCMC algorithm}
The pMCMC algorithm used in this work is given in Algorithm 1 for an observed dataset $\mathbf{y} = \{y_k\,|\,k=1,2,\dots,T\}$.
We use a Metropolis-Hastings MCMC algorithm \citep{metropolis1953equation}, proposing new parameters using a proposal distribution $q(. | \theta)$.
In step 6 of Algorithm 1, we need to evaluate the likelihood, $p(\mathbf{y}|\theta)$, in calculating the acceptance probability for the proposed move.
We replace the likelihood with an unbiased estimate of the likelihood, $\hat{p}(\mathbf{y}|\theta)$, obtained from a particle filter. 
% ---------------------------------------------------------------------------------------------

\begin{algorithm}[h!]
%\floatname{algorithm}{Algorithm 1}
%\renewcommand{\thealgorithm}{}
\caption{Particle MCMC}
\label{protocol1}
\begin{algorithmic}[1]
\STATE Initialise parameters, $\theta_0$.
\STATE Run a particle filter (see Algorithm 2) to compute an estimate of the marginalised likelihood $\hat{p}(\mathbf{y} | \theta_0)$,
where $\mathbf{y}$ is the observed data.
\FOR{$j = 1:N$}
\STATE Draw parameters $\theta^*$ from a proposal distribution $q(. | \theta_{j-1})$. \\
\STATE Run a particle filter to compute an estimate of the marginalised likelihood $\hat{p}(\mathbf{y} | \theta^*)$. \\
\STATE Accept the proposed move with probability $\alpha$ where
$$ \alpha = \text{max}\left(1,\frac{\pi(\theta^*)q(\theta_{j-1} | \theta^*)\hat{p}(\mathbf{y} | \theta^*) }
{\pi(\theta_{j-1})q(\theta^* | \theta_{j-1})\hat{p}(\mathbf{y} | \theta_{j-1})}\right).$$
If the move is accepted, set $\theta_j = \theta^*$, otherwise set $\theta_j = \theta_{j-1}$.
\ENDFOR
\end{algorithmic}
\end{algorithm}

% ---------------------------------------------------------------------------------------------

\begin{algorithm}[h!]
%\floatname{algorithm}{Algorithm 1}
%\renewcommand{\thealgorithm}{}
\caption{Bootstrap particle filter }
\label{protocol1}
\begin{algorithmic}[1]
\STATE Sample a collection of particles $\{x_1^1, \dots, x_1^M \}$ from an initial density $p(x_1)$. \\
\FOR{$i =1:M$}
\STATE Compute the weights for each particle, $i$, via $w_1^i = p(y_1|x_1^i,\theta).$\\
\STATE Find the normalised weights $$\tilde{w}_1^i = \frac{w_1^i}{\sum_{j=1}^M w_1^j}.$$
\ENDFOR
\STATE Resample $N$ times with replacement from the collection of particles $\{x_1^1, \dots, x_1^M \}$
with probabilities given by the normalised weights $\{w_1^1, \dots, w_1^M \}$.   
\FOR{$t = 1:(T-1)$}
\FOR{$i =1:M$}
\STATE Evolve the current collection of particles according to the forward model,
by drawing $x_{t+1}^i \sim p(x_{t+1}| x_t^i,\theta).$
\STATE Compute the weights for each particle, $i$, via $w_{t+1}^i = p(y_{t+1}|x_{t+1}^i,\theta).$\\
\STATE Find the normalised weights $$\tilde{w}_{t+1}^i = \frac{w_{t+1}^i}{\sum_{j=1}^M w_{t+1}^j}.$$
\ENDFOR
\STATE Resample $M$ times with replacement from the collection of particles $\{x_{t+1}^1, \dots, x_{t+1}^M \}$
with probabilities given by the normalised weights $\{\tilde{w}_{t+1}^1, \dots, \tilde{w}_{t+1}^M \}$.   
\ENDFOR
\STATE Obtain an estimate of the marginal likelihood using the unnormalised weights, via 
$$\hat{p}(\mathbf{y} | \theta) = \prod_{t=1}^T \frac{1}{M} \sum_{i=1}^{M}w_t^i .$$
\RETURN $\hat{p}(\mathbf{y} | \theta)$ 
\end{algorithmic}
\end{algorithm}

% ---------------------------------------------------------------------------------------------

Details of the bootstrap particle filter algorithm \citep{gordon1993novel} used within the pMCMC algorithm are given in Algorithm 2. A particle filter represents the state of the system via a population of weighted particles \citep{doucet2000sequential}. We obtain an estimate of the likelihood by successively updating the hidden state of the system (represented via the particles) and comparing this hidden state with the observed data at each observed time point, to give new weights for the particles according to how well they match the observed data. To prevent a degenerate situation where the state of the system is represented solely by a single particle, it is necessary to resample from the population of particles according to their weights.

% ---------------------------------------------------------------------------------------------
\subsection*{Implementation of Markov chain Monte Carlo methods}

We use a Metropolis-Hastings algorithm to run the MCMC algorithm with a bootstrap particle filter \citep{gordon1993novel} using 400 particles to provide an estimate of the likelihood. 
As a proposal distribution, we use the kernel $K(.,\theta) \sim N(\theta,\Sigma)$,
where 
\[ \Sigma = \begin{bmatrix}
    0.5 & 0 \\
    0 & 0.05
\end{bmatrix}. \] 
We run the Markov chain for $N=50,000$ steps with thinning of $m=2$ which gives a minimum effective sample size of $n_{\text{eff}}=743$ for the slowest chain to converge, corresponding to the smallest value of $\Delta t$.
We demonstrate the convergence of our Markov chains in Supplementary Figure \ref{Fig:traceplots}.
We choose the number of particles in the filter by considering the variance in estimating the log likelihood using the particle filter at the true values of the parameters used to generate the synthetic data, and balancing this with the time needed to run the particle filter to obtain a single estimate.
The variance and computational cost are shown in Supplementary Figure \ref{Fig:particles}.
To further tune the number of particles for optimal efficiency of the pMCMC algorithm, the recommendations in \citet{sherlock2015efficiency} and \citet{pitt2012some} could be employed.
The prior is uniform on the log of the parameters over the intervals $[-1.70,1.30]$ and $[-5,1]$ for $\lambda$ and $\sigma$, respectively.

\imdonenow

% ---------------------------------------------------------------------------------------------
\beginresponse
\subsection*{Approximate Bayesian computation}

An alternative method commonly used for parameter estimation for models with intractable likelihoods is approximate Bayesian computation, or ABC \citep{pritchard1999population,beaumont2002approximate}.
Suppose we can simulate data from a generative model, $x \sim g(x | \theta)$. An example would be simulating a path from our VJP model. To generate samples from the posterior distribution, we repeatedly generate parameters from the prior, $\theta \sim \pi(\theta)$, and use these parameters in our model to simulate synthetic data, $x \sim g(x | \theta)$. We compare the simulated data with the true observed data and, if it is similar enough, we accept the sampled parameter as a sample from the posterior. In cases where the data are discrete, we can consider whether simulated data, $x$, are equal to observed data, $y$, and accept parameters correspondingly,
which gives exact samples from the posterior. Often, though, our models provide continuous data and the acceptance rate from an exact comparison is prohibitively small. 
Instead we can take a distance function and consider when the distance between $x$ and $y$ is within a chosen tolerance $\epsilon$,
that is $d(x,y) < \epsilon$.
This introduces an approximation which becomes exact only in the limit $\epsilon \rightarrow 0$.

In practice, datasets can be high dimensional, resulting in very low acceptance rates for simulated data.
By using summary statistics of data instead of the full datasets, we can reduce the dimensionality and obtain higher acceptance rates.
Using (insufficient) summary statistics introduces another approximation into the method, meaning we compare $d(s(x),s(y)) < \epsilon $, where $s(x)$ gives the summary statistics for dataset $x$.
We present the ABC method in Algorithm 3, and will show a comparison between application of pMCMC and ABC for parameter estimation of a VJP model based on time series data. 

% ---------------------------------------------------------------------------------------------

\begin{algorithm}[h!]
\floatname{algorithm}{Algorithm 3}
\renewcommand{\thealgorithm}{}
\caption{ABC rejection sampling}
\label{protocol1}
\begin{algorithmic}[1]
\STATE Sample parameter value $\theta$ from the prior $\pi(\theta)$.
\STATE Generate synthetic dataset from the model $x \sim g(x | \theta)$.
\STATE Accept sample $\theta$ if $d(s(x),s(y)) < \epsilon$,
for a distance function, $d(\cdot,\cdot)$, summary statistics, $s(\cdot)$, and tolerance $\epsilon$.
\end{algorithmic}
\end{algorithm}

In practice, the choice of summary statistics can have a strong effect on the efficiency of the ABC algorithm.
Here, we perform ABC with three different choices of summary statistic.
First, we consider using the full data, a vector of the noisy observed angle changes, $z_i$, $i=1,\dots ,T$.
Second, we use a simple threshold-based summary statistic which produces a count of the number of observed angle changes with magnitude above a certain threshold, $h$.
That is $s(\mathbf{z}) = \sum_{i=1}^T \mathds{1}_{|z_i| > h}$.
This provides an intuitive one-dimensional summary statistic. 
Finally, we use a transition matrix as a summary statistic of the data, an approach described by \citet{jones2015inference}.
A transition matrix allows us to summarise time series data via a two-dimensional binning of the observed angle changes, based on the current observed angle change and the previous observed angle change.
We bin the observed angle change time series data into an $n$ by $n$ matrix, where $n=5$.
This summary statistic provides a much lower dimensional summary of the data for small values of $\Delta t$, although for large values of $\Delta t$ we may end up with higher dimensional (sparse) data compared to the full time series data, as in this case there will be fewer than $n^2$ observations.
The distance function, $d(s(x),s(y))$, used also plays a role in the parameter estimation possible via ABC, and has been investigated in other work \citep{jones2015inference,prangle2015adapting,harrison2017automatic,bernton2017inference}. 

\beginresponse In using Algorithm 3, we generate $N=1,000,000$ \imdonenow synthetic datasets based on parameters sampled from the prior, calculate the Euclidean distance between these synthetic datasets and our observed data, and select the parameters corresponding to the $0.1\%$ of the datasets closest to the observed data.
As for the pMCMC approach, we take a duration of time series $T_{\text{final}}=64\,\text{s}$, and a prior uniform on the log of the parameters on the intervals $[-1.70,1.30]$ and $[-5,1]$ for $\lambda$ and $\sigma$, respectively.

% ---------------------------------------------------------------------------------------------

\section*{Results}

We are able to investigate the effects of experimental design such as the discretisation, $\Delta t$, of a time series on the estimated biophysical parameters, using the framework for inferring the parameters of a VJP described in the Section `Methods'. We demonstrate the effects of restrictions imposed by experimental constraints on a trade-off between discretisation in time versus measurement noise. Our results can be used to provide guidance for experimental design choices.
\beginresponse
As we vary experimental design hyperparameters, such as $\Delta t$, we will produce a separate posterior distribution for each of the model parameters, $\lambda$ and $\sigma$, for each value of the hyperparameter.
This approach allows us to illustrate directly the effects of the experimental design hyperparameter on the inferred posterior distributions.
\imdonenow

% ---------------------------------------------------------------------------------------------

\subsection*{Increasing $\Delta t$ results in an abrupt breakdown in the posterior}\label{fixed_sigma}

\begin{figure*}[h!]
\begin{center}
\includegraphics[width=\textwidth]{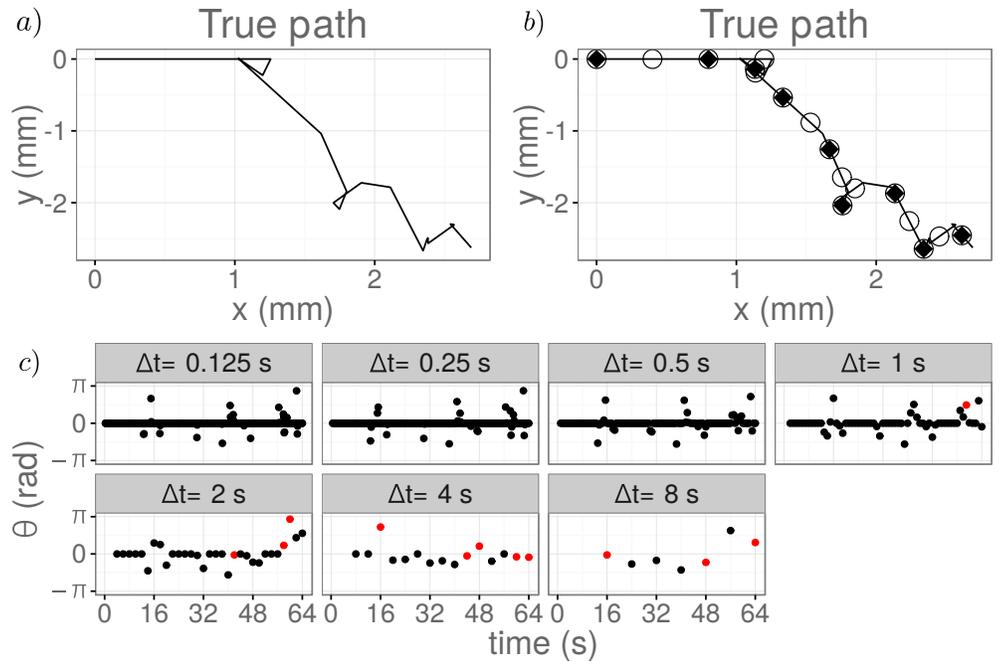}
\end{center}
\caption{The data used in the pMCMC inference discretised at differing resolutions.
The true trajectory is shown without discretisation in a).
The same path is shown in b) with circle markers to show the observed positions with $\Delta t=4\,\text{s}$ and filled triangular markers to show observed positions with $\Delta t=8\,\text{s}$.
The corresponding observed angle changes are shown in c) for different values of $\Delta t$.
Observations corresponding to multiple reorientations in a single time interval are highlighted in red.
Parameters used in generating these data were a reorientation rate of $\lambda = 0.2\,\text{s}^{-1}$, a run speed of $c=50\,\mu \text{m} \text{s}^{-1}$
and a total duration of observation of $T_{\text{final}}=64\,\text{s}$.
A circular uniform distribution was used for the reorientation kernel. 
Observations are shown without measurement noise.
}
\label{Fig:datasets}
\end{figure*}

We first assume that the magnitude of the noise on the observed angle, $\sigma$, is fixed.
We vary the sampling frequency used to collect the data (corresponding to using different values of $\Delta t$).
We generate observed (\textit{in silico}) data by simulating one single trajectory directly from the VJP model.
We discretise this trajectory to generate observed angle changes with a temporal resolution of $\Delta t$, and add independent Gaussian noise with zero mean and variance $\sigma^2$ to these observations, to represent measurement noise.
Using datasets generated from the same simulated path discretised at different temporal resolutions, $\Delta t$, as shown in Fig \ref{Fig:datasets}, we infer the posterior distribution for the reorientation rate, $\lambda$, of the VJP model and the magnitude of the measurement noise, $\sigma$, via pMCMC.

% ---------------------------------------------------------------------------------------------

% ---------------------------------------------------------------------------------------------

\begin{figure}[htbp]
\begin{center}
\includegraphics[width=\columnwidth]{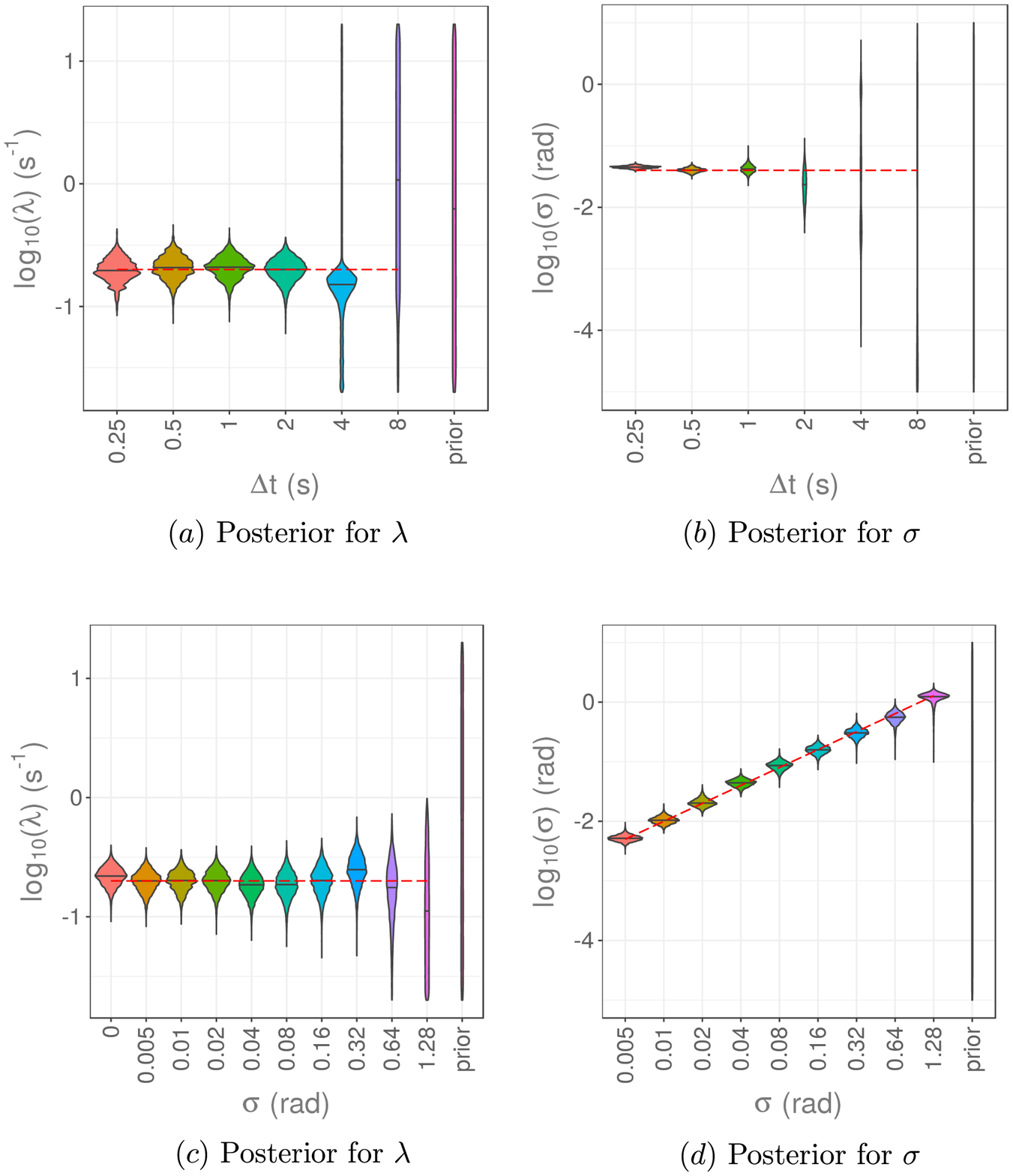}
\end{center}
\caption{Results of parameter estimation via pMCMC for the reorientation rate, $\lambda$, and measurement noise, $\sigma$.
Data collected with different values of $\Delta t$ and fixed measurement noise, $\sigma = 0.04\,\text{rad}$, were used in a) and b).
Data generated with a fixed value of $\Delta t = 1\,\text{s}$ and different values of the measurement noise, $\sigma$, were used in c) and d).
The marginal posterior distributions for $\lambda$ are given in a) and c), while the marginal posterior distributions for $\sigma$ are shown in b) and d).
The red dashed lines indicate the true values of parameters used in simulation of the datasets.
\beginsecondresponse
A reorientation rate of $\lambda = 0.2\,\text{s}^{-1}$ was used throughout.
\imdonenow
%The central rectangle of each boxplot shows the quartiles of the plotted distribution, with the central line giving the median. The whiskers above and below the rectangle show the maximum and minimum of the distribution unless there are outliers beyond 1.5 times the interquartile range below or above the 1st or 3rd quartiles, which are shown as dots.
}
\label{Fig:results_dt_sig}
\end{figure}

The results of using pMCMC to infer model parameters are shown in Fig \ref{Fig:results_dt_sig} for $\Delta t = 1/8, 1/4,1/2, 1, 2, 4, 8$ seconds, with noise of unknown magnitude $\sigma$, where $\sigma = 0.04$ rad.
\beginresponse The estimated posterior distributions are very similar for small $\Delta t$, but we observe an abrupt breakdown in the quality of the posterior distributions obtained as $\Delta t$ is increased. \imdonenow
This breakdown in posterior quality arises for values of $\Delta t \ge 2 \,\text{s}$, as multiple reorientation events in a time interval start to become more common (see also Fig \ref{Fig:events}).
Provided that $\Delta t$ and $\lambda$ are such that the probability of multiple reorientations in a time interval is small, we obtain accurate estimates of the joint posterior distribution for the model parameters.

% ---------------------------------------------------------------------------------------------

% ---------------------------------------------------------------------------------------------

\subsection*{Increasing $\sigma$ increases the variance of the posterior for $\lambda$}

To investigate sensitivity of the posterior to measurement noise, we now fix the discretisation, $\Delta t$, and vary the measurement noise amplitude, $\sigma$, used to create each dataset.
Applying the same analysis as in the previous subsection, for a fixed value of \beginresponse $\Delta t = 1\,\text{s}$, \imdonenow
we obtain posterior distributions for the reorientation rate, $\lambda$, and measurement noise, $\sigma$, as shown in Fig \ref{Fig:results_dt_sig}. 
We find that increasing the noise amplitude, $\sigma$, increases the variance in the posterior distribution obtained for the reorientation rate, $\lambda$.
In addition, we are able to accurately estimate the value of $\sigma$ used to generate the datasets.

We note that the presence of noise in the datasets results in a bias towards smaller estimates of the reorientation rate, $\lambda$.
This can be explained intuitively by reasoning that some reorientations leading to very small observed angle changes are mistaken for noise as $\sigma$ increases.
%which can be seen by comparison with the case with no measurement noise ($\sigma = 0$).

% ---------------------------------------------------------------------------------------------

\subsection*{More data provides better estimates}

\begin{figure}[h!]
\begin{center}
\includegraphics[width=0.5\columnwidth]{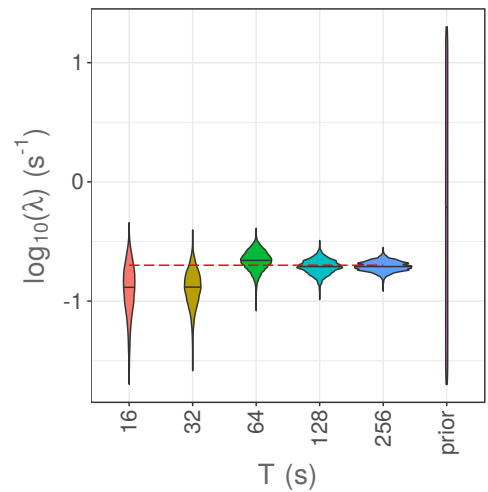}
\end{center}
\caption{Results of parameter estimation via pMCMC for the reorientation rate, $\lambda$, with $\Delta t =1\,\text{s}$ and $\sigma = 0\,\text{rad}$, whilst varying $T_{\text{final}}$ to give different sized datasets. 
}
\label{Fig:results_T}
\end{figure}

Let $T_{\text{final}}$ be the total duration of our observations of the system. 
For a fixed value of the time discretisation, $\Delta t$, varying $T_{\text{final}}$ is equivalent to gathering a bigger dataset.
We fix $\Delta t = 1\,\text{s}$ and allow $T_{\text{final}}$ to vary so that we can consider the effects of collecting more data.
\beginresponse In practice, collecting more data experimentally may come at a cost.
Quantifying the benefits of collecting more data with an \textit{in silico} model can aid decisions about whether it is worthwhile to collect a larger dataset. \imdonenow
Posteriors for the reorientation rate, $\lambda$, are shown in Fig \ref{Fig:results_T}.

% ---------------------------------------------------------------------------------------------

% ---------------------------------------------------------------------------------------------

It is clear that larger datasets result in less bias and less variance for estimates of the reorientation rate, $\lambda$. 
However, we note that running the particle filter within the pMCMC algorithm becomes much more computationally intensive as the size of the dataset increases,
scaling as $\mathcal{O}(T)$ as the size of the dataset increases.

% ---------------------------------------------------------------------------------------------

\subsection*{Experimental constraints} \label{photon_budget}
%An experimentalist may be interested in implications of this inference for experimental design.
We investigate the implications of these results for experimental design by considering an imaging experiment to observe the position of a particle of interest (for example, a bacterium) at regular time points. 
The behaviour of our system of interest happens over a certain timescale inherent to the biological process, so we fix the total duration for the imaging experiment, $T_{\text{final}}$, based on this timescale.
We consider how best to choose the time between successive observations, $\Delta t$, given the restriction of a fixed photon budget.
That is, we assume that the biological sample can only be exposed to a fixed number of photons before phototoxicity or photobleaching significantly reduce the quality of, or destroy, any further potential data.
\beginsecondresponse
We assume that the sample is only exposed to photons during imaging.
The results of \citet{zhao2011photon} suggest that the signal to noise ratio (SNR) is proportional to the square root of the time between successive frames, $\Delta t$, giving a relationship of the form
\begin{equation}
 \text{SNR} = \kappa \sqrt{\Delta t},
\end{equation}
where $\kappa$ is a constant that depends on the imaging set up, but not other experimental design choices.
\imdonenow
Assuming that for our model the noise is of magnitude $\sigma$, we find an inverse square relationship between $\sigma$ and $\Delta t$, such that
\begin{equation} \label{inv_sq_root}
 \sigma = \frac{K}{\sqrt{\Delta t}},
\end{equation}
for a new constant $K$ which is $\kappa$ times the average angle change.

% ---------------------------------------------------------------------------------------------

\begin{figure}[h!]
\begin{center}
\includegraphics[width=\columnwidth]{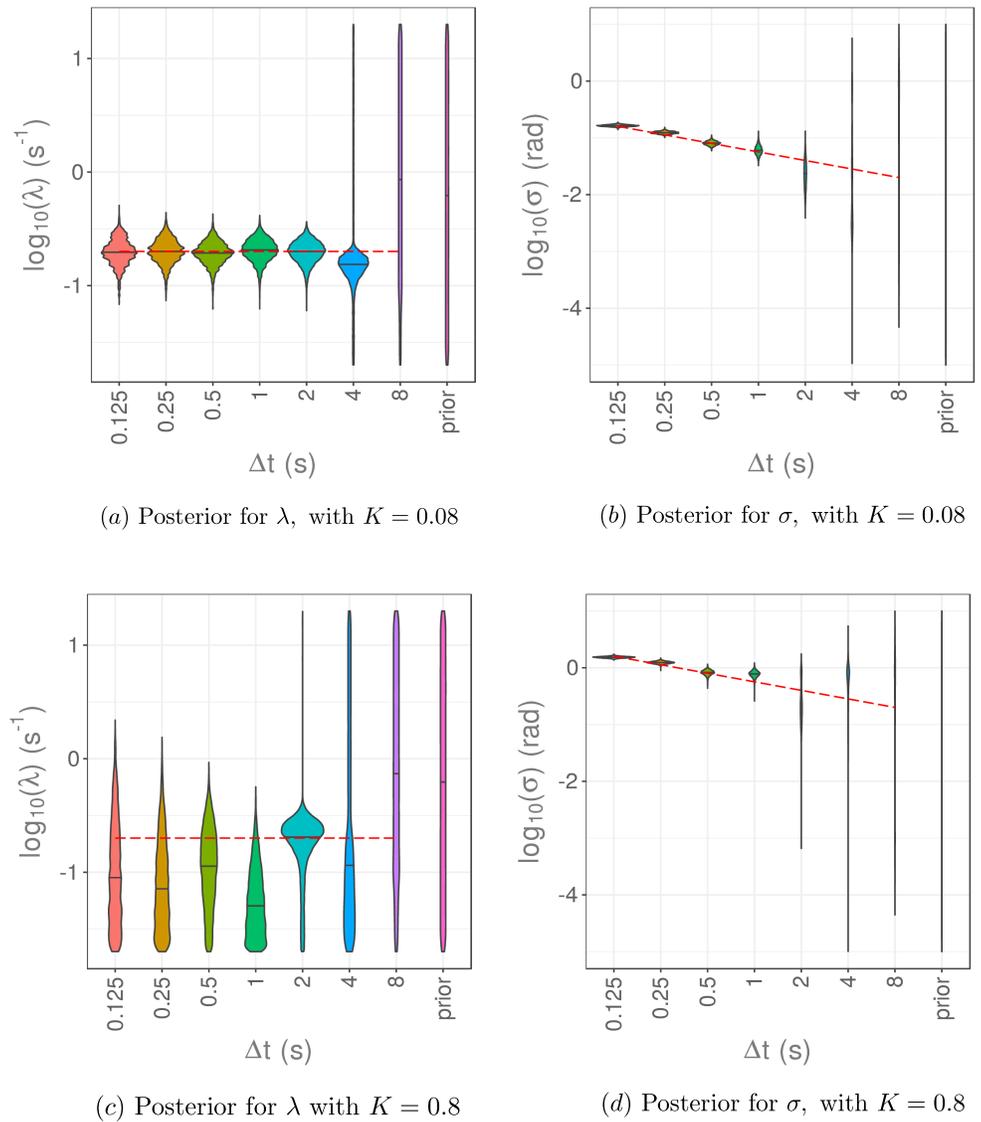}
\end{center}
\caption{Results of parameter estimation via pMCMC for the reorientation rate, $\lambda$, and measurement noise, $\sigma$,
using data generated with a fixed value of $\Delta t$ and different values of the measurement noise, $\sigma$.
We have varied both $\sigma$ and $\Delta t$ with an inverse square root relationship between these, as in Eq.\eqref{inv_sq_root}.
The posterior distributions for $\lambda$ are given in a) and c), while the posterior distributions for $\sigma$ are shown in b) and d).
The proportionality constant is $K=0.08$ for a) and b), and $K=0.8$ for c) and d).
}
\label{Fig:results_sqrt}
\end{figure}

% ---------------------------------------------------------------------------------------------

To investigate how to choose $\Delta t$ given a fixed photon budget, we set a value of the proportionality constant and vary $\sigma$ and $\Delta t$ according to this relationship.
We take proportionality constants $K = 0.08$ and $K=0.8$ in Fig \ref{Fig:results_sqrt}.
A larger value of the proportionality constant, $K$, corresponds to worse imaging conditions, in that for a fixed value of $\Delta t$, the noise in the images obtained will be greater.
Therefore we expect our inference method to perform worse for a larger value of $K$. 
We then ask the question, for a given value of $K$, how should we choose $\Delta t$ to improve the parameter estimation process? 
Our results in Fig \ref{Fig:results_sqrt} suggest that the value of $\Delta t$ should be taken as small as possible, even if this increases the noise present in the data.

% ---------------------------------------------------------------------------------------------

\subsection*{Comparison to approximate Bayesian computation} \label{compare_abc}

A common approach for mathematical and computational models where the likelihood is intractable is to apply ABC for parameter inference \citep{pritchard1999population,beaumont2002approximate}.
Although ABC produces samples from an approximate posterior, rather than the exact posterior distribution, it is intuitively simple to understand and implement.
ABC methods have been applied to parameter estimation for biased, persistent random walk models very similar to our VJP model \citep{liepe2012calibrating,taylor2013p38,jones2015inference,weavers2016systems}.
For these methods, the choice of which summary statistics to use has notable effects on the approximate posterior distributions obtained via ABC.
\beginresponse We consider three different summary statistic choices: the full time series of observed angle changes, a threshold-based summary statistic, and a transition matrix summary statistic. \imdonenow

\begin{figure}[h!]
\begin{center}
\includegraphics[width=0.85\columnwidth]{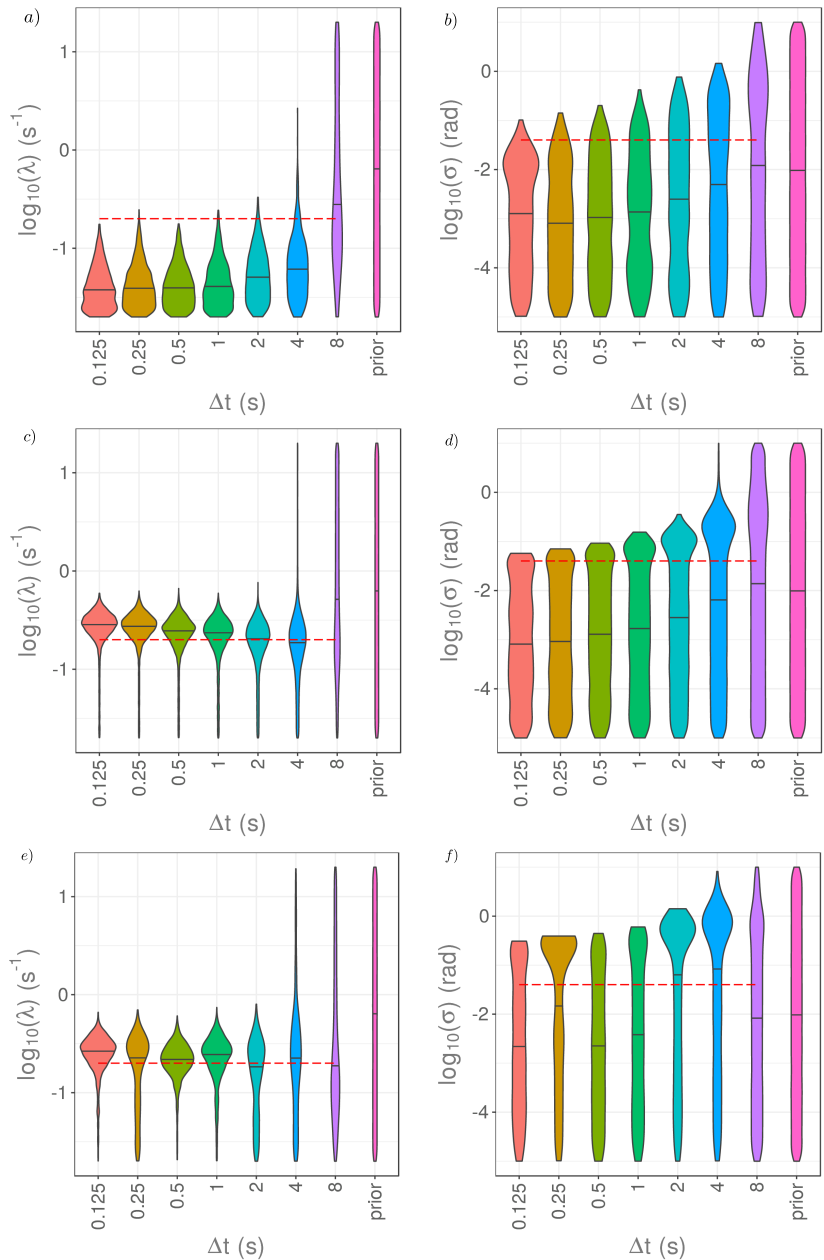}
\end{center}
\caption{Parameter estimation via ABC rejection sampling for the reorientation rate, $\lambda$, in a), c), and e) and for the measurement noise, $\sigma$, in b), d), and f)
using \beginresponse different summary statistics \imdonenow and data collected with noise of magnitude $\sigma = 0.04\,\text{rad}$, for different values of $\Delta t$. 
\beginresponse $N=1,000,000$ \imdonenow datasets were generated and parameter samples corresponding to the closest $0.1\%$ of the datasets were retained to give the approximate posterior.
\beginresponse As summary statistics, we take the full time series of observed angle changes in a) and b).
In c) and d), we use a simple threshold summary statistic counting the number of observed angle changes with magnitude above a threshold, here taken as $h=0.1\, \text{rad}$.
In d) and e), we use a transition matrix summary statistic using $n=5$ bins to discretise the observed angle change.
\imdonenow
}
\label{Fig:ABC_ts}
\end{figure}

We compare the quality of resulting posterior distributions obtainable with ABC to those from pMCMC.
Our results, shown in Fig \ref{Fig:ABC_ts},
\beginresponse
 indicate that the choice of summary statistic has a substantial effect on the inferred posterior distribution.
 When the dimensionality of the data observed is high, ABC performs poorly at approximating the posterior for the parameters of our VJP, which we were able to sample exactly using pMCMC.
 For the full time series summary statistic, the dimensionality is high (up to 511 dimensions for $\Delta t=0.125\,\text{s}$), meaning that the approximation we obtain to the posterior is very poor (see Fig \ref{Fig:ABC_ts}a) and b)).
 Since reorientation events are rare, particularly for small $\Delta t$, a greater fraction of the distance between simulated datasets and observed data is accounted for by the observation noise.
 ABC excludes part of the parameter space considered in the prior, but does poorly at identifying the reorientation rate, $\lambda$, since reorientations are rare events.

The threshold summary statistic is effective at identifying the number of reorientation events and provides a reasonable approximation of the marginal posterior for $\lambda$.
However, it offers very little information about the observation noise, $\sigma$, except in relation to the threshold chosen.
As a result the approximate marginal posteriors for $\sigma$ are poor (see Fig \ref{Fig:ABC_ts}c) and d)).
Another summary statistic more informative about $\sigma$ could be used in addition here to improve results.

% ---------------------------------------------------------------------------------------------

% ---------------------------------------------------------------------------------------------

%We consider also the effects on the approximate posteriors sampled via ABC of using summary statistics, such as transition matrices \citep{jones2015inference}, which enable us to reduce the dimensionality of the data. 
When using the transition matrix summary statistic, the approximate posteriors for small $\Delta t$ provide better estimates of the true reorientation rate and are much closer to the true posterior, as shown in Fig \ref{Fig:ABC_ts}e) and f).
The transition matrix is able to capture the distribution of the angle changes and dependence on recent history, whilst also reducing the dimensionality of the data.
However, the transition matrix is unable to distinguish angle changes smaller than the resolution of the discretisation, $2\pi / n$, and so offers limited information about the measurement noise, $\sigma$.
This issue could be mitigated by increasing the number of bins used, $n$, to give a finer discretisation of the observed angle changes (which would require sufficient data), or by targeting the noise with an additional summary statistic.

We still notice a deterioration in the quality of the approximate posterior distributions as $\Delta t$ increases when using ABC for each choice of summary statistic even though there are no explicit assumptions made about multiple reorientations within a time interval, as was the case when using pMCMC.
This suggests that the lack of information content about the parameters may be a property of the data, rather than the inference method.
For large values of $\Delta t$, the data contains limited information about the model parameters, particularly the reorientation rate, $\lambda$. 
\imdonenow

% ---------------------------------------------------------------------------------------------

To obtain potentially improved results for inference with ABC for this type of data (time series observations of angle changes), we could consider a more systematic choice of summary statistics \citep{fearnhead2012constructing,barnes2012considerate,blum2013comparative} or a more efficient version of the ABC algorithm, such as ABC-SMC \citep{sisson2007sequential,toni2009approximate},
which \beginresponse applies sequential Monte Carlo methods to generate a sequence of approximations to the posterior distribution, using the previous posterior distribution to propose parameters for the next approximation. \imdonenow
Other improvements could be possible by applying a regression adjustment, via linear regression \citep{beaumont2002approximate} or using nonlinear regression techniques such as with a neural network \citep{blum2010nonlinear}. 

% ---------------------------------------------------------------------------------------------
\beginresponse
\subsection*{Model misspecification}

In general, in applying a model to a real world dataset, any model that we choose will be an approximation of the true data generating process.
Before applying our inference methods to real world data, where we will fit to a model that is but an approximation to the true biological process, it is important to check the robustness of our method to fit a misspecified model.
We investigate this robustness computationally by considering a misspecified model which should be no longer misspecified in an appropriate limit.
It has previously been shown by \citet{frazier2017model} that inference with ABC on a misspecified model can concentrate posterior mass around different pseudo-true parameter values depending on the version of ABC used. 

\begin{figure}[h!]
\begin{center}
\includegraphics[width=\columnwidth]{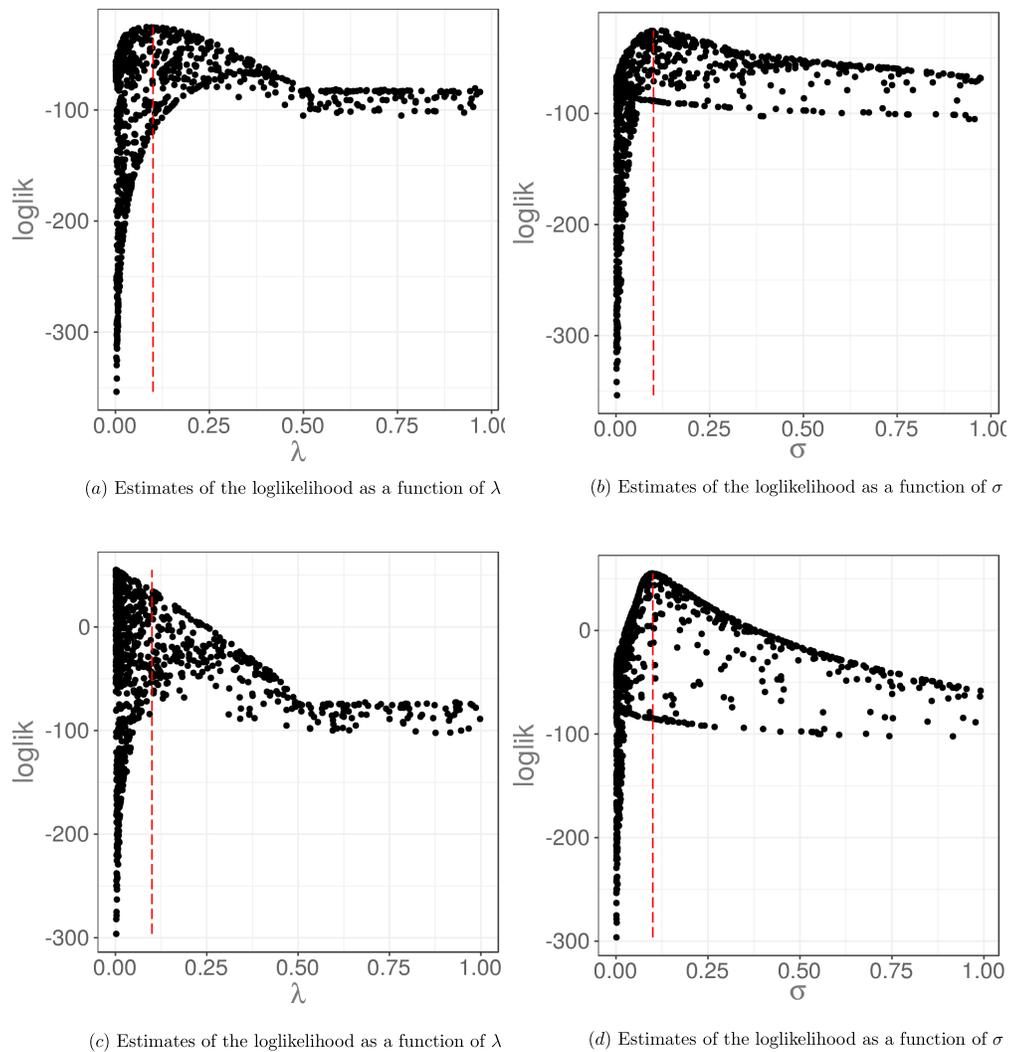}
\end{center}
\caption{\beginresponse
Estimates of the log likelihood via a particle filter for data generated with the true model and under a misspecified model.
In a) and b), we show estimates of the log likelihood for different values of $\lambda$ and $\sigma$ based on a dataset simulated with a uniform reorientation kernel, and assuming this same reorientation kernel in estimating the log likelihood via the particle filter.
The peak of the log likelihood coincides with the true values of the parameters shown by the red dashed line.
In c) and d), we show estimates of the log likelihood for different values of $\lambda$ and $\sigma$ based on a dataset simulated with a wrapped normal reorientation kernel with dispersion $\gamma=0.1$.
However, we assume a uniform reorientation kernel in the particle filter.
This model misspecification results in a slight shift in the peak of the log likelihood compared to the true parameter values used shown by the red dashed line. 
\imdonenow
\beginsecondresponse
The log likelihood estimates will be biased in the case of the misspecified model.
\imdonenow
}
\label{Fig:misspecification}
\end{figure}

Here, we demonstrate the robustness of our inference framework, using the pMCMC algorithm within a hidden states framework, to give relevant estimates for a misspecified model. 
We generate synthetic datasets using a wrapped normal reorientation kernel with dispersion parameter $\gamma$, but choose to estimate the model parameters with a model different to the data generating process by assuming a uniform reorientation kernel.
As previously, we attempt to infer posterior distributions for the reorientation rate, $\lambda$, and the measurement noise, $\sigma$. 
The effect of misspecifying the model will be that some of the transmission probabilities used in the particle filter estimate of the likelihood of model parameters given the data will be wrong (see supplementary Figure \ref{Fig:misspecification_transition_probs}) This will give rise to some bias in our estimates of the likelihood within the particle filter (as demonstrated by Fig. \ref{Fig:misspecification}), and hence to some bias in our final approximation of the posterior.
We will consider the effect of varying the dispersion parameter, $\gamma$, in the reorientation kernel used to generate the synthetic dataset, on the resulting posterior distribution for the model parameters.
In the limit $\gamma \rightarrow \infty$, the wrapped normal distribution converges to the uniform distribution on $(-\pi, \pi]$, meaning that the model is no longer misspecified.
The misspecification of the model is most pronounced for smaller values of $\gamma$.

%--------------------------------------------------------------------------------

%--------------------------------------------------------------------------------------

\begin{figure*}[p]
\begin{center}
\includegraphics[width=0.85\columnwidth]{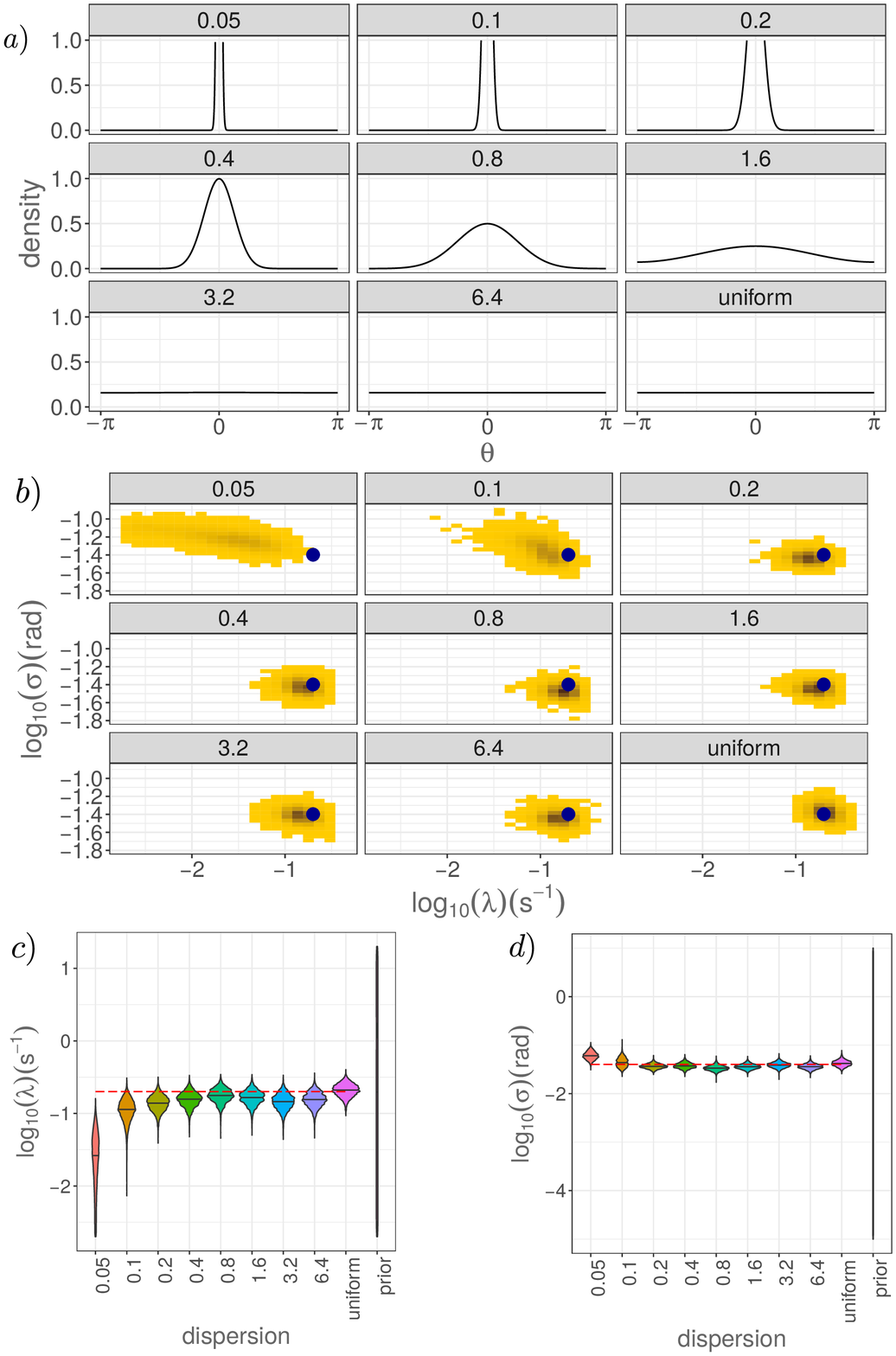}
\end{center}
\caption{ \beginresponse
The posterior distribution obtained under model misspecification converges to the distribution obtained with the true model as the misspecified model approaches the true model.
We assume a wrapped normal reorientation kernel with dispersion parameter $\gamma$ and vary this dispersion parameter.
For large $\gamma$, this tends in distribution towards a uniform reorientation kernel, as shown in a).
The posterior distributions obtained from performing inference with data generated for different values of $\gamma$ are shown in b),
with the corresponding marginal distributions shown in c) and d).
In b), the true parameter values used are shown by the dark blue circle.
In c) and d), the true parameter values are given by the red dashed line.
Parameters used to generate the synthetic datasets are $\lambda = 0.2\,\text{s}^{-1},\sigma=0.04\,\text{rad},\Delta t=0.25\,\text{s},c=50\mu \text{m}\text{s}^{-1}$, and the details of the pMCMC are as in Fig \ref{Fig:results_dt_sig}.
\imdonenow
}
\label{Fig:misspecification_2d_posterior}
\end{figure*}

We find, for a fixed value of the measurement noise, $\sigma$, that a reasonable approximation to the true posterior can be obtained for values of the dispersion, $\gamma$, larger than the measurement noise, $\sigma$.
The approximate posterior distributions obtained are shown in Fig. \ref{Fig:misspecification_2d_posterior}.
When the dispersion, $\gamma$, is a similar magnitude to the measurement noise, which here is $\sigma = 0.04\, \text{rad}$, then reorientation events are frequently missed resulting in low estimates of the reorientation rate.
For larger values of the dispersion parameter, we are able to robustly sample from the posterior distribution for $\lambda$ and $\sigma$, despite notable differences in the misspecified reorientation kernel compared to the assumed uniform reorientation kernel.
In this case, when a reorientation occurs, the observed angle changes are sufficiently different to the background measurement noise, that often it is interpreted as a reorientation event by the particle filter, despite error in the emission probabilities due to the model misspecification. 

% ---------------------------------------------------------------------------------------------

\subsection*{Application to RNA transport dataset}

To demonstrate the effectiveness of our inference framework, we apply it here to a dataset of tracks obtained from imaging the transport of RNA-protein (RNP) complexes in a \textit{Drosophila} oocyte.
These complexes move on microtubules via molecular motors \citep{parton2014subcellular}. 
Occasionally, the complexes fall off a microtubule, and reattach on a different microtubule moving in a different direction.
We neglect the diffusive stationary phase between falling off and reattaching on microtubules,
meaning that we assume a running phase is followed immediately by another running phase.
We apply our modelling framework, assuming an exponentially distributed running time distribution, $f_{\tau}(t) = \lambda \exp{(-\lambda t)} $, with constant running speed, $c$, and a uniform reorientation kernel, $f_{\Phi}(\phi) = 1/2\pi \, \mathds{1}_{(-\pi, \pi]}(\phi) $. 

\begin{figure}[h!]
\begin{center}
\includegraphics[width=0.9\columnwidth]{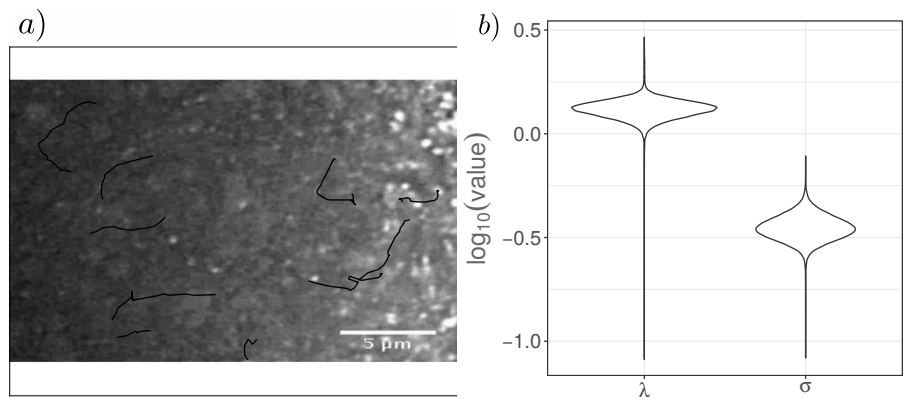}
\end{center}
\caption{\beginresponse
Posterior distribution for model parameters $\lambda$ and $\sigma$ for RNA transport in a \textit{Drosophila} oocyte.
The results are based on 10 tracks labelled manually and sampled at a resolution of $\Delta t = 0.478 \,\text{s}$, with subposteriors obtained for each track and combined to give a single posterior via Consensus Monte Carlo.
The tracks are shown drawn on the first frame in a), although not all tracks start from the same time point.
The marginal posterior distributions for the reorientation rate, $\lambda$, and the measurement noise, $\sigma$, are shown in b). 
\imdonenow
}
\label{Fig:rna_posteriors}
\end{figure}

We take 10 tracks of separate complexes moving in \textit{Drosophila} nurse cells obtained from an \textit{in vivo} imaging dataset of movements of staufen protein available in \citet{zimyanin2008vivo}.
Each track is short since the dataset is imaged in a single plane in the $z$ direction, meaning that complexes move out of the frame of view frequently.
The tracks used are shown in Fig. \ref{Fig:rna_posteriors}a).
We infer a subposterior distribution for each individual track and combine these together to produce a single posterior distribution combining data from different tracks.
To combine the subposteriors, we use the consensus Monte Carlo algorithm of \citet{scott2016bayes} developed for running MCMC on large datasets, via the implementation in the R \citep{R2013language} package \texttt{parallelMCMCcombine} \citep{miroshnikov2014parallelmcmccombine}.
This produces an approximate posterior distribution for the turning rate, $\lambda$, and measurement noise, $\sigma$, as shown in Fig. \ref{Fig:rna_posteriors}b).
We find $95\%$ credible intervals for the turning rate $\lambda$ of $[1.06,\, 1.58]\, \text{s}^{-1}$ and for the measurement noise $\sigma$ of $[0.27,\, 0.47] \, \text{rad}$, respectively.
Based on a VJP model corresponding to Brownian motion, the relation $\lambda = c^2 / (2D)$ holds for a constant running speed $c$ and diffusion constant $D$. 
The running speed in active transport for staufen RNPs is of the order $0.5 \, \mu\text{ms}^{-1}$ \citep{zimyanin2008vivo},
and based on the size of the staufen protein, we can estimate its diffusion coefficient as $1 \, \mu\text{m}^2\text{s}^{-1}$ \citep{kumar2010mobility}, which gives a similar order of magnitude for $\lambda$ as the estimate obtained here.
Evidently the dataset used here is very noisy, but nonetheless we are able to obtain estimates for the turning rate in a model of RNA transport.

\imdonenow
% ---------------------------------------------------------------------------------------------

\section*{Discussion}

In this work, we have considered parameter estimation of a VJP model for biological transport and how insights from this parameter inference process can inform experimental design. 
We generated estimates of the posterior distribution for parameters of a VJP model based on noisy datasets collected at varying temporal resolutions. 
To perform this parameter inference, we used pMCMC and derived the appropriate emission probabilities.
We observed an abrupt breakdown in the quality of the posteriors obtained when \beginresponse decreasing the temporal resolution. \imdonenow
This transition corresponds to a breakdown in modelling assumptions underlying the derivation of the emission probabilities.
For example, as $\Delta t$ increases we see multiple reorientations in a time interval.
\beginresponse These assumptions are necessary in the development of our inference framework which relies on hidden states consisting of binary variables indicating whether or not a reorientation occurred in a time interval.
%By using the binary hidden variables, $X_k$, we are making an approximation and assuming reorientations are rare. This assumption is needed for the purposes of simplification to allow us to make progress here.
\imdonenow

Increasing the magnitude of the noise in the data slowly decreased the quality of the median posterior estimates and increased the posterior variance.
In general, better estimates were obtained when more data was available, either by decreasing $\Delta t$,
or by increasing the total duration of the experimental observations, $T_{\text{final}}$.
\beginresponse These results are intuitive, but the real benefit here is in quantifying the effects of choices of these experimental hyperparameters to enable researchers to make decisions about experimental design. \imdonenow

We also compared parameter estimation with pMCMC to that with ABC, and results suggest that different methods are appropriate in different situations, dependent on the data available.
In particular, pMCMC performs well provided the assumptions made in deriving the emission probabilities hold, which is true for small $\Delta t$. 
\beginsecondresponse For larger values of $\Delta t$, neither ABC nor pMCMC provide accurate samples from the full posterior distribution.
The information about the model parameters present in the data for large $\Delta t$ is limited.
In terms of computational cost comparisons between ABC and pMCMC, an ABC rejection sampler is highly parallelisable over independent parameter samples \citep{owen2015scalable} whereas pMCMC is less amenable to simple parallelisation.
Making use of this parallelisation, our implementation of ABC rejection sampling was more computationally efficient than pMCMC.
We show a quantitative comparison between the methods in Supplementary Figure \ref{Fig:computational_cost}.
We note that the computational cost is strongly problem and implementation dependent. 
\imdonenow

Through comparison between inference with pMCMC and ABC, our results highlight a weakness of inference with ABC,
in that it performs poorly for high dimensional data, and also a weakness of our application of pMCMC, which is that it relies upon assumptions about the number of reorientation events in a time step; when this fails our posterior estimates are no longer accurate.
We note, additionally, that pMCMC allows us to sample from the exact posterior for parameters from a model (given that model is appropriate to describe the data),
whereas via ABC we obtain approximate posteriors for a fixed tolerance, $\epsilon$, which only become exact in the limit as $\epsilon \rightarrow 0$.

In Section `Methods', we described a framework for inference using pMCMC, and made certain assumptions about the VJP model,
such as a separation in timescales between runs and reorientations, and a memoryless exponential distribution for the running time.
Although many of these are standard assumptions, it would be possible to perform the same analysis in a more general model. 
For instance, if a running time distribution was chosen that does not satisfy the memoryless property, we could introduce an extra hidden variable, $s$, for the time since the last reorientation.
In addition, in this work we have described parameter estimation via a hidden states formulation of the VJP model using dependence on hidden states from two time intervals.
This can be extended to hidden states from three time intervals, which can allow rare consecutive reorientation events
to be handled more accurately, albeit at greater computational cost. 

Given a fixed photon budget and a trade-off between temporal sampling frequency and measurement noise, our results in the Section `Experimental constraints'
indicate that a small value of $\Delta t$ should be used for the discretisation in time i.e. that motile individuals should be imaged as frequently as possible.
In practice, there may be disadvantages to this choice of $\Delta t$.
Computationally, conducting parameter inference via pMCMC will be significantly more expensive, although the computational run time may still be small in comparison to the duration of an experimental protocol.
Datasets with higher noise present may also be much harder to interpret. 
Here, we have assumed that the noise present in the data is applied to the observed angle change. 
In reality there is noise on each pixel, which may contribute to an uncertainty in identifying the observed position of the object of interest.

\beginresponse
Additionally, we considered using our particle MCMC inference framework to fit data from a misspecified model; our results indicate that the framework is robust to moderate amounts of model misspecification. This allowed us to be confident in applying our framework to a real dataset of RNA tracks to consider the motility of RNA-protein complexes moving on the cytoskeleton, demonstrating the ability of our framework to obtain parameter estimates in challenging conditions with very noisy data. 
\imdonenow

% ---------------------------------------------------------------------------------------------

\section*{Acknowledgements}
% This work was supported by funding
% from the Engineering and Physical Sciences Research Council
% (EPSRC) (grant no. EP/G03706X/1).
% REB is a Royal Society Wolfson Research Merit Award holder,
% and would like to thank the Leverhulme Trust for a Leverhulme Research Fellowship.
The authors are grateful to Richard Parton and Ilan Davis for helpful discussions of the application to RNA transport.

%---------------------------------------------------------------------------------------------------

\section*{Supporting information}

\paragraph*{S1 Appendix.}
\label{S1_Appendix}
{\bf Comparison with simulations.} Description of simulations to verify the analytic form for the emission probabilities.

\paragraph*{S2 Fig.}
\label{S2_Fig}
{\bf Comparison with simulations.} Comparison of the analytic form for the emission probabilities with results from simulated paths. Results are shown for hidden states of the form 0,1 for (A), 1,0 in (B), and 1,1 in (C).

\paragraph*{S3 Fig.}
\label{S3_Fig}
{\bf Probability of hidden state sequences.} Probability of sequences of hidden states as $\Delta t$ varies, showing where multiple reorientations in a time interval appear as $\Delta t$ increases.

\beginresponse
\paragraph*{S4 Fig.}
\label{S4_Fig}
{\bf Emission probabilities for misspecified model.} Comparison between simulated distributions of observed angle changes using a misspecified reorientation kernel and the theoretical emission probabilities assumed, based on a uniform reorientation kernel. Dispersion parameter, $\gamma$, varies over rows in the figure as indicated in the panels, with $\gamma=0.4$ for (A,B,C), $\gamma=1.0$ for (D,E,F), and $\gamma=6.4$ for (G,H,I). The pattern of hidden states varies across columns of the figure as indicated in the panels, with states 0,1 for (A,D,G), 1,0 for (B,E,H), and 1,1 for (C,F,I). 

\paragraph*{S5 Fig.}
\label{S5_Fig}
{\bf Traceplots and autocorrelation functions for MCMC chains.} Traceplots and autocorrelation functions are shown for one of the MCMC chains simulated as a convergence diagnostic.
\imdonenow

\paragraph*{S6 Fig.}
\label{S6_Fig}
{\bf Analysis of number of particles in the particle filter.} The variance of estimates of the log-likelihood obtained via the particle filter depends on the number of particles used in the particle filter. The number of particles also affects the computational cost of the particle filter algorithm. (A) shows how the distribution of log-likelihood estimates changes as the number of particles in the particle filter varies. (B) shows how the computation time and variance change as the number of particles varies.

\beginsecondresponse
\paragraph*{S7 Fig.}
\label{S7_Fig}
{\bf The computational cost of pMCMC and ABC per effective sample size.} Comparison of the computational cost of parameter estimation with pMCMC and ABC per effective sample size, showing also the scaling of this cost with $\Delta t$.
\imdonenow

\paragraph*{S8 Code}
\label{S8_Code}
{\bf Code implementing Bayesian inference via pMCMC for a VJP model.} Example code to implement pMCMC for a VJP model, as described in Section ``Methods'', is available in a .zip folder or at $\href{https://github.com/shug3502/pmmc_inference_for_vjps}{\tt{https://github.com/shug3502/pmmc\_inference\_for\_vjps}}$

\pagebreak

% ---------------------------------------------------------------------------------------------

\beginsupplement
% \section*{S1 Appendix}
% 
% % ---------------------------------------------------------------------------------------------
% 
% \subsection*{Comparison with simulations} \label{Comparison}
% 
% To verify our emission probabilities calculated in the Section `Derivation of emission probabilities', 
% we can compare the distributions of observed angle changes from simulations with those predicted theoretically.
% We simulate from the VJP model for a known pattern of hidden states and assume no measurement noise is present. Performing these checks for hidden states of the form $0, 1$, we obtain results as shown in Fig \ref{Fig:checks}a); there is excellent agreement between simulation and theory which verifies our theoretical results.
% For hidden states of the form $1, 0$, we must condition on the value of the previous angle change. 
% We demonstrate agreement between the theory and simulations, for different values of the previous angle change, in Fig \ref{Fig:checks}b).
% Similarly, we show corresponding results for hidden states of the form $1, 1$ in Fig \ref{Fig:checks}c). 

% ---------------------------------------------------------------------------------------------

\begin{figure}[htbp!]
\begin{center}
\includegraphics[width=0.7\columnwidth]{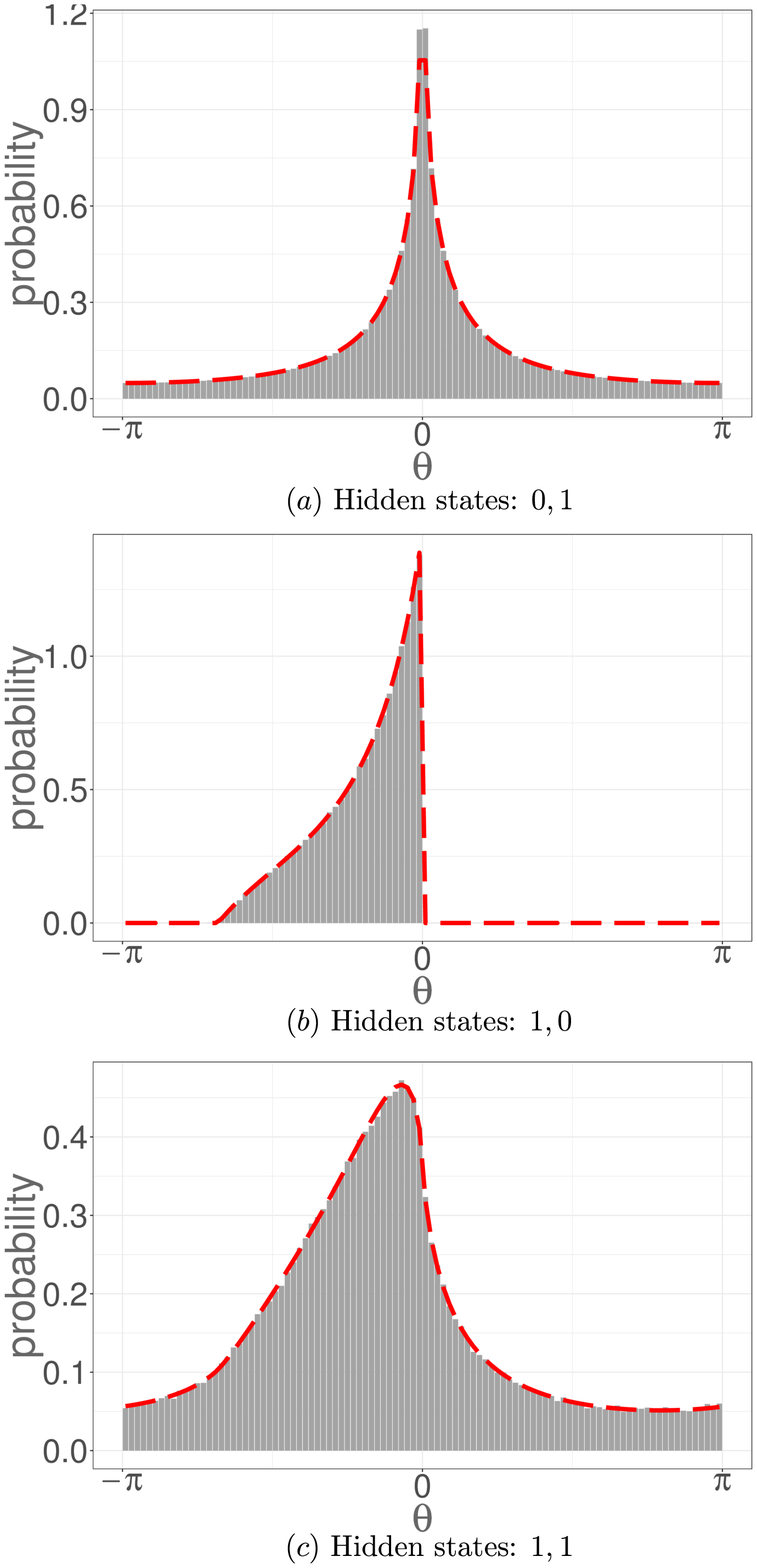}
\end{center}
\caption{Comparison between simulated results and theoretical predictions of the observed angle change for hidden states of the form $0, 1$ in a), $1, 0$ in b),
and $1, 1$ in c). The simulated results are shown by the histogram and the theoretical prediction for the observed angle change distribution is shown as the red dashed line. 
For both b) and c), we have conditioned on an observed angle change in the previous time interval of $0.1$ rad,
and for c) we also conditioned on an observed angle change prior to that of $-1.0$ rad. 
To generate these results, we used $N=10^7$ simulated trajectories with running speed $c = 50\,\mu \text{m}\text{s}^{-1}$, uniform reorientation kernel, reorientation rate $\lambda = 0.2\,\text{s}^{-1}$ and time discretization $\Delta t = 1\,\text{s}$.
}
\label{Fig:checks}
\end{figure}
\clearpage

% ---------------------------------------------------------------------------------------------

\begin{figure}[h!]
\begin{center}
\includegraphics[width=\columnwidth]{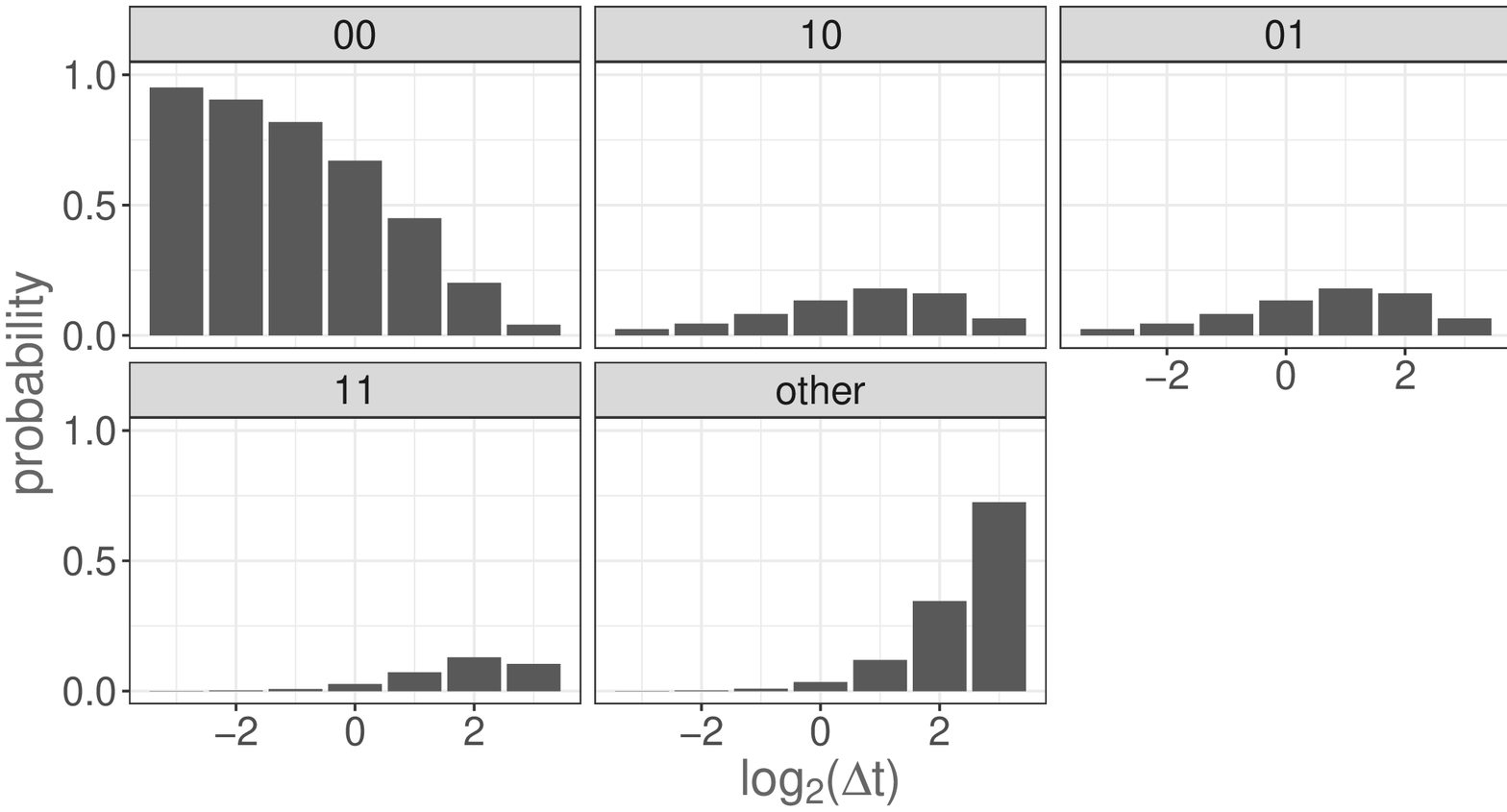}
\end{center}
\caption{The probability of sequences of hidden states as $\Delta t$ varies with reorientation rate $\lambda = 0.2\,\text{s}^{-1}$.
For large values of $\Delta t$, the assumptions of the model start to break down as multiple reorientations appear within a single time interval.}
\label{Fig:events}
\end{figure}
\clearpage

% ---------------------------------------------------------------------------------------------

\begin{figure}[h!]
\begin{center}
\includegraphics[width=\columnwidth]{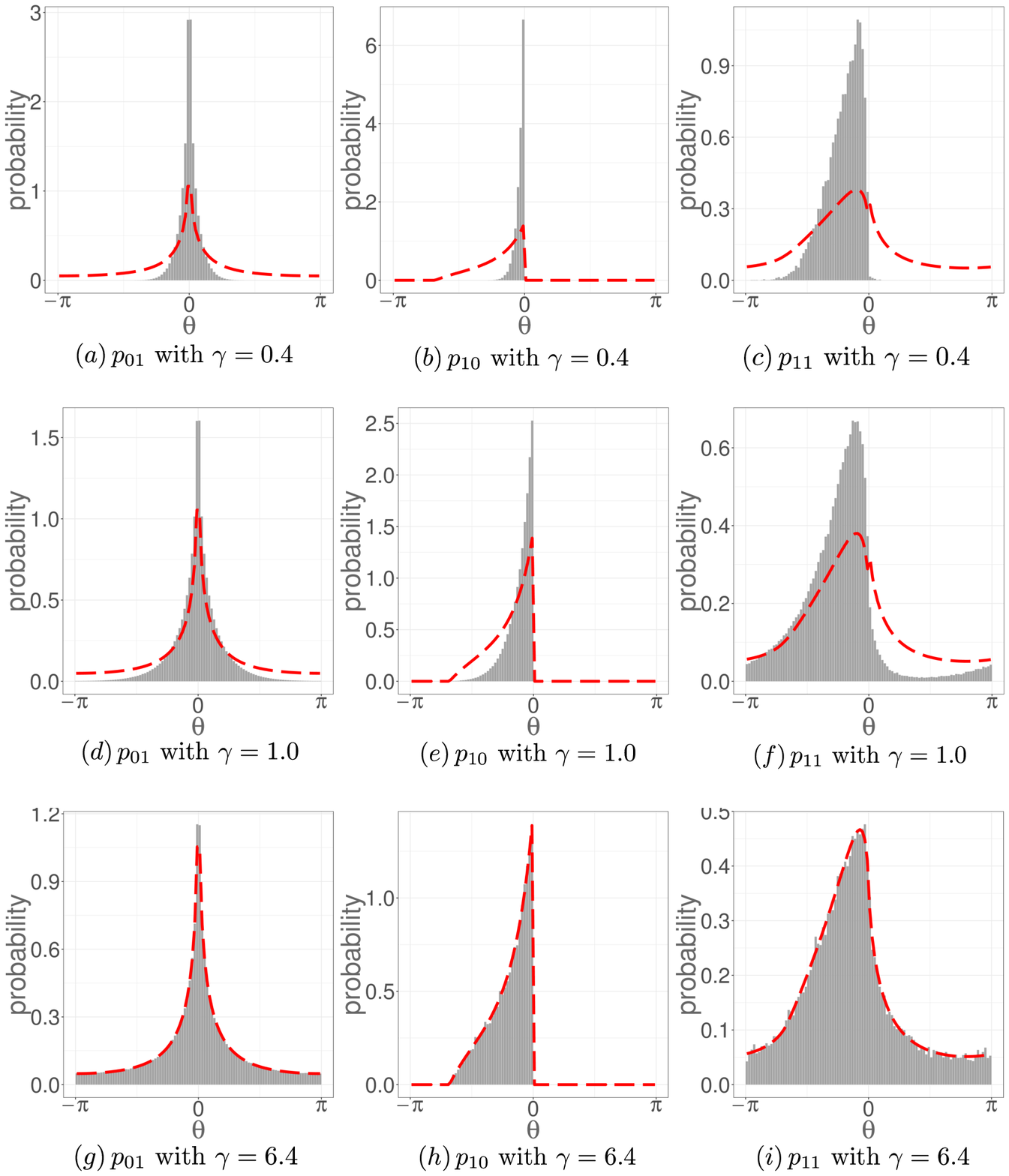}
\end{center}
\caption{\beginresponse Comparison between assumed distribution of observed angle changes and simulated distributions when using a misspecified model for the reorientation kernel.
We simulate $N=10^7$ angle changes using a wrapped normal reorientation kernel with dispersion parameter $\gamma$ given a certain sequence of hidden states (0,1 in a), d), g); 1,0 in b), e), h); 1,1 in c), f), i)) and show a grey histogram of the simulated observed angle changes.
To demonstrate the misspecification in the emission probabilities, we plot the assumed theoretical distribution of the observed angle changes as a red dashed line, based on assuming that the reorientation kernel is a uniform distribution.
The model is more misspecified for a smaller value of the dispersion parameter $\gamma$.
As in Fig \ref{Fig:checks}, we have conditioned on an observed angle change in the previous time interval of $0.1$ rad for b), c), e), f), h), and i).
For c), f), and i), we also conditioned on an observed angle change prior to that of $-1.0$ rad. 
We used a dispersion parameter, $\gamma$, in the reorientation kernel of $\gamma = 0.4$ for a), b), and c), $\gamma=1$ for d), e), and f), and $\gamma=6.4$ for g), h), and i).
\imdonenow
}
\label{Fig:misspecification_transition_probs}
\end{figure}
\clearpage

%---------------------------------------------------------------------------------------------------

\begin{figure}[h!]
\begin{center}
\includegraphics[width=\columnwidth]{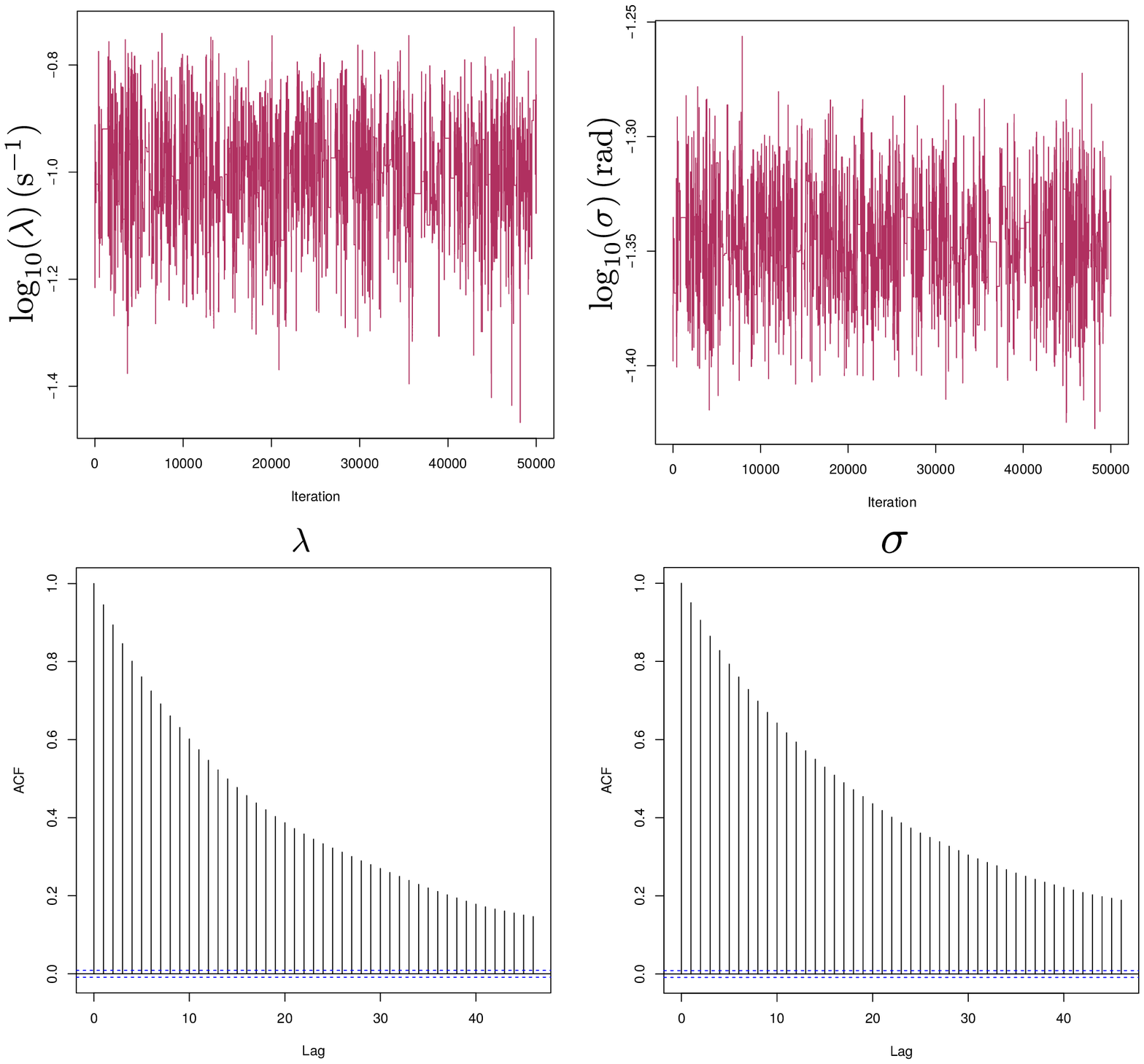}
\end{center}
\caption{ \beginresponse
Traceplots and autocorrelation functions for MCMC chains to analyse convergence are shown for data generated with parameters $\lambda=0.2\,\text{s}^{-1}$,$\Delta t = 0.25\,\text{s}$, $\sigma=0.04\,\text{rad}$, as for the first posterior shown in Fig \ref{Fig:results_dt_sig}a).
Similar results are seen in sampling for other posterior distributions shown.
\imdonenow
}
\label{Fig:traceplots}
\end{figure}
\clearpage

%--------------------------------------------------------------------------------------------------

\begin{figure}[h!]
\begin{center}
\includegraphics[width=\columnwidth]{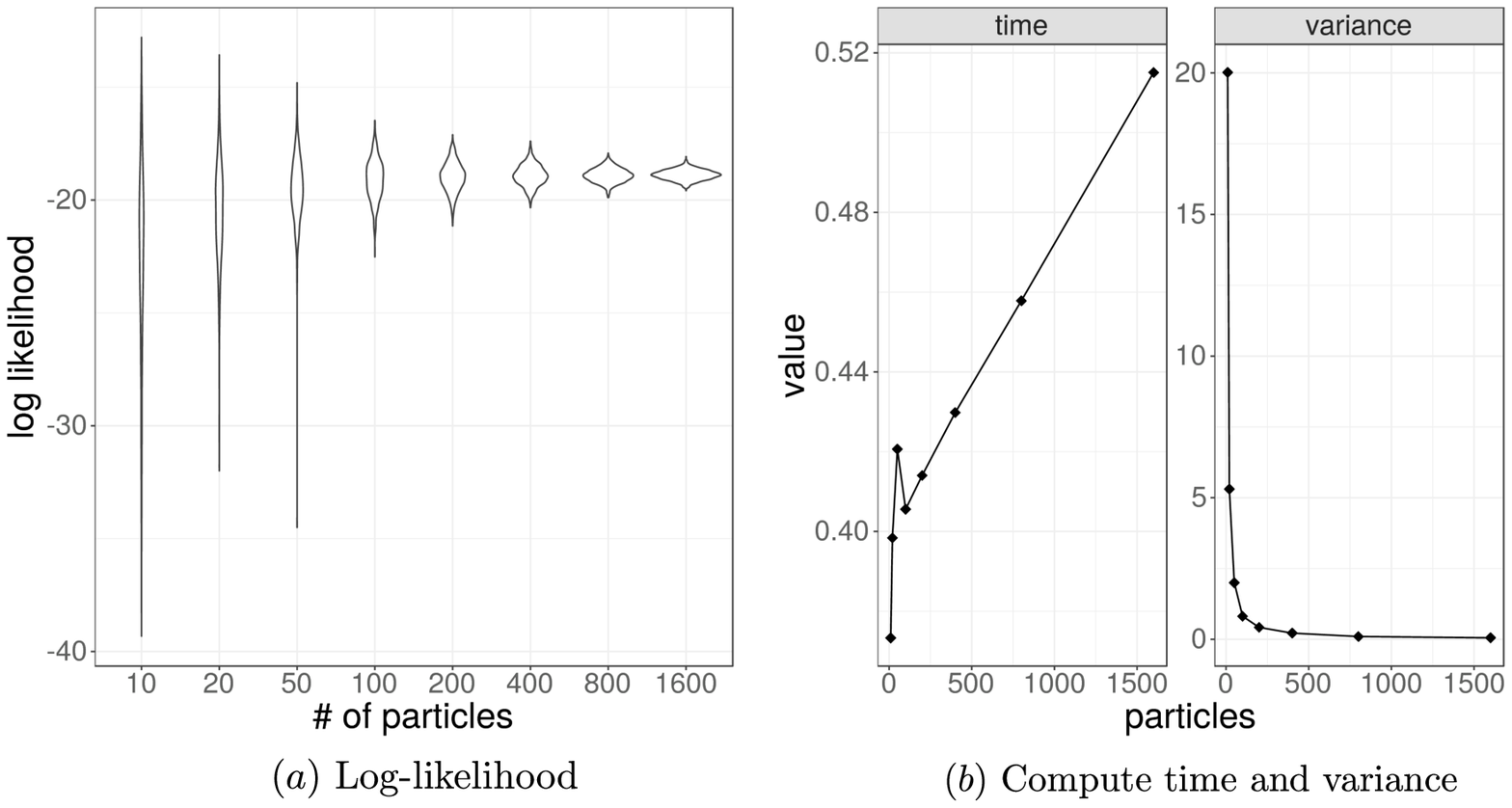}
\end{center}
\caption{ \beginresponse
The number of particles used in the particle filter affects the variance of estimates of the log-likelihood.
However, a higher computational cost is required to use more particles.
In a), we show how the distribution of the log-likelihood estimates varies (provided the filter does not become degenerate) as we change the number of particles.
We use 1000 runs of the particle filter with the specified number of particles and estimate the log-likelihood at the true value of the parameters ($\lambda=0.2\,\text{s}$ and $\sigma=0.08\,\text{rad}$) used to generate a synthetic dataset.
In b), we illustrate the variance in the (nondegenerate) log-likelihood estimates, and the mean time to obtain a single estimate.
A moderate increase in the compute time to run the particle filter offers substantial decrease in the variance of the log-likelihood estimates.
To strike a reasonable balance, we use $400$ particles to generate the results presented in this work.  
\imdonenow
}
\label{Fig:particles}
\end{figure}
\clearpage

\begin{figure}[h!]
\begin{center}
\includegraphics[width=\columnwidth]{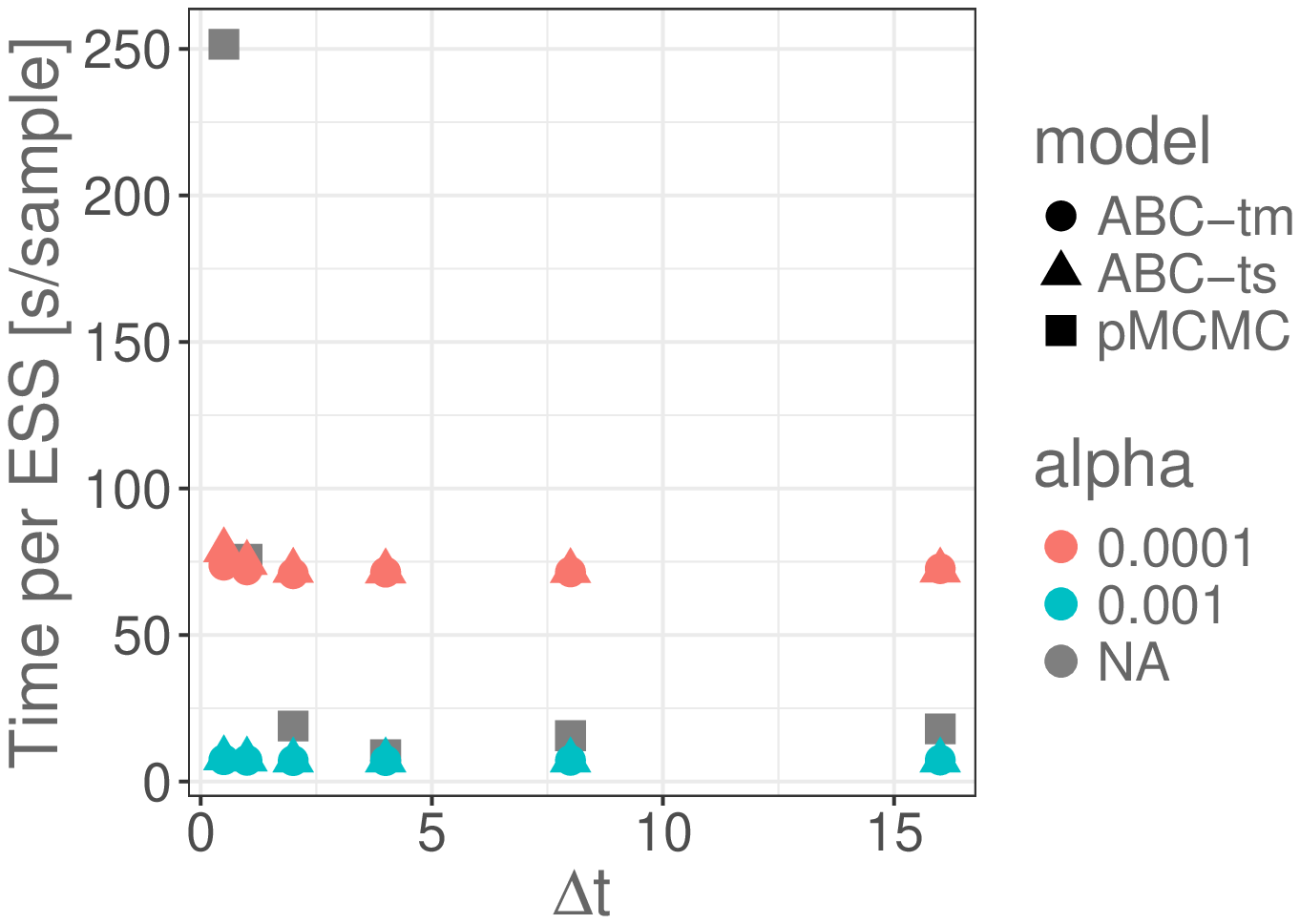}
\end{center}
\caption{ \beginsecondresponse
The computational cost of parameter estimation with pMCMC and ABC per effective sample size depends on $\Delta t$.
We quantify here a comparison between the computational cost of the pMCMC and ABC methods for parameter estimation when varying $\Delta t$.
Datasets were generated with $\sigma = 0.04\,\text{rad}$ as in Fig. \ref{Fig:results_dt_sig}a).
The cost to produce a sample via pMCMC increases as $\Delta t$ decreases, as shown by the grey squares.
The cost for ABC remains approximately constant with $\Delta t$ as we simulate data from the model a fixed number of times, $N$.
The acceptance rate, $\alpha$, in ABC affects the sample size we produce for a fixed number of simulations, $N$.
For the results in Fig. \ref{Fig:ABC_ts}, an acceptance rate of $\alpha=0.1\%$ was used. This gives a cost per sample lower than for pMCMC, shown by the blue circles and triangles.
For a smaller acceptance rate, $\alpha=0.01\%$ (shown by red circles and triangles), the cost per sample is much higher and the plots of the posterior are unchanged compared to those for $\alpha=0.1\%$. 
The results for pMCMC are given by squares, ABC with transition matrix summary statistics are shown as circles and ABC with time series summary statistics are shown as triangles.
We note that the computational cost depends strongly on the problem considered and the implementation used.
\imdonenow
}
\label{Fig:computational_cost}
\end{figure}
\clearpage

% ---------------------------------------------------------------------------------------------

\nolinenumbers

%
%\bibliography{/home/harrison/Documents/Summary_stats/Latex/MT16.bib}
%\bibliographystyle{plos2015}
%\bibliographystyle{unsrtnat} %{plainnat}

% Compile your BiBTeX database using our plos2015.bst
% style file and paste the contents of your .bbl file
% here. See http://journals.plos.org/plosone/s/latex for 
% step-by-step instructions.
% 

\end{document}